\documentclass[longauth]{aa}  

\usepackage{arydshln}
\usepackage{graphicx}
\usepackage{todonotes}
\usepackage{txfonts}
\usepackage{mathtools}
\usepackage{color}

\begin{document} 

   \title{The extended atmosphere and circumstellar environment of the cool evolved star VX~Sagittarii as seen by MATISSE\thanks{Based on the observations made with VLTI-ESO Paranal, Chile under the programme IDs 0103.D-0153(D, E, G). The data are available at oidb.jmmc.fr}}

\titlerunning{The cool evolved star VX Sagittarii}
%   \subtitle{I. Overviewing the $\kappa$-mechanism

   \author{A. Chiavassa
          \inst{1,2,3}
          \and
          K. Kravchenko\inst{4}
          \and
          M. Montarg\`{e}s\inst{5}
          \and
          F. Millour\inst{1}
          \and
          A. Matter\inst{1}
          \and
          B. Freytag\inst{6}
          \and
          M. Wittkowski\inst{2}
          \and 
          V. Hocd\'e \inst{7} 
          \and 
          P. Cruzal\`ebes \inst{1}
          \and F. Allouche \inst{1}
          \and B. Lopez \inst{1}
          \and S.~Lagarde \inst{1}
          \and R. G. Petrov \inst{1}
          \and A. Meilland \inst{1}
          \and S. Robbe-Dubois \inst{1}
          \and K.-H. Hofmann \inst{8}
          \and G. Weigelt \inst{8}
          \and P. Berio \inst{1}
          \and P. Bendjoya \inst{1}
          \and F.~Bettonvil \inst{9} 
          \and A.~Domiciano de Souza \inst{1}
          \and M.~Heininger \inst{8}
          \and Th. Henning \inst{10}
          \and J. W. Isbell \inst{10}
          \and W. Jaffe \inst{9}
          \and L.~Labadie \inst{11}
          \and M.~Lehmitz \inst{10} 
          \and K. Meisenheimer \inst{10}
          \and A.~Soulain \inst{12}
          \and J.~Varga \inst{9,13}
          \and J.-C.~Augereau \inst{12}
          \and R.~van~Boekel \inst{10}
          \and L.~Burtscher \inst{9}
          \and W.~C. Danchi \inst{14}
          \and C. Dominik \inst{15}
          \and J. Drevon \inst{1}
          \and V.~G\'amez~Rosas \inst{9}
          \and M.R. Hogerheijde \inst{9,15}
          \and J. Hron \inst{16}
          \and L.~Klarmann \inst{10}
          \and E.~Kokoulina \inst{1}
          \and E.~Lagadec \inst{1}
          \and J. Leftley \inst{1}
          \and L.~Mosoni \inst{17}
          \and N.~Nardetto \inst{1}
          \and C.~Paladini \inst{18}
          \and E.~Pantin \inst{19} 
          \and D.~Schertl \inst{8}
          \and P.~Stee \inst{1}
          \and L.~Szabados \inst{13,20} 
          \and R.~Waters \inst{21,22}
          \and S. Wolf \inst{23}
          \and G.~Yoffe \inst{9}
}

   \institute{Universit\'e C\^ote d'Azur, Observatoire de la C\^ote d'Azur, CNRS, Lagrange, CS 34229, Nice,  France \\
                \email{andrea.chiavassa@oca.eu}
                \and
   European Southern Observatory, Karl-Schwarzschild-Str. 2, 85748 Garching, Germany
   \and
   Max-Planck-Institut f\"{u}r Astrophysik, Karl-Schwarzschild-Stra\ss{}e 1, 85741 Garching, Germany
   \and
   Max Planck Institute for extraterrestrial Physics, Giessenbachstra\ss{}e 1, 85748 Garching, Germany
   \and
   LESIA, Observatoire de Paris, Universit\'e PSL, CNRS, Sorbonne Universit\'e, Universit\'e de Paris, 5 place Jules Janssen, 92195 Meudon, France
   \and
   Theoretical Astrophysics, Department of Physics and Astronomy at Uppsala University, Regementsv\"agen 1, Box 516, SE-75120
Uppsala, Sweden
   \and
   Nicolaus Copernicus Astronomical Centre, Polish Academy of Sciences, Bartycka 18, 00-716 Warszawa, Poland
   \and
   Max-Planck-Institut f\"ur Radioastronomie, Auf dem H\"ugel 69, D-53121 Bonn, Germany
   \and 
   Leiden Observatory, Leiden University, Niels Bohrweg 2, NL-2333 CA Leiden, the Netherlands
    \and   
   Max Planck Institute for Astronomy, K\"onigstuhl 17, D-69117 Heidelberg, Germany
   \and
I. Physikalisches Institut, Universit\"at zu K\"oln, Z\"ulpicher Str. 77, 50937, K\"oln, Germany
   \and
   Univ. Grenoble Alpes, CNRS, IPAG, 38000, Grenoble, France
   \and
    Konkoly Observatory, Research Centre for Astronomy and Earth Sciences, E\"otv\"os Lor\'and Research Network (ELKH), Konkoly-Thege Mikl\'os \'ut 15-17, H-1121 Budapest, Hungary
\and
NASA Goddard Space Flight Center, Astrophysics Division, Greenbelt, MD 20771, USA
\and
Anton Pannekoek Institute for Astronomy, University of Amsterdam, Science Park 904, 1090 GE Amsterdam, The Netherlands
\and
Department of Astrophysics, University of Vienna, 1180 Vienna,T\"urkenschanzstrasse 17, Austria
\and
Zselic Park of Stars, 064/2 hrsz., 7477 Zselickisfalud, Hungary
\and
European Southern Observatory, Alonso de Cordova 3107, Casilla 19, Santiago 19001, Chile
\and
AIM, CEA, CNRS, Universit\'e Paris-Saclay, Universit\'e Paris
Diderot, Sorbonne Paris Cit\'e, F-91191 Gif-sur-Yvette, France
\and
CSFK Lend\"ulet Near-Field Cosmology Research Group, Budapest,
Hungary
\and
Institute for Mathematics, Astrophysics and Particle Physics, Radboud University, P.O. Box 9010, MC 62 NL-6500 GL Nijmegen, the
Netherlands
\and
SRON Netherlands Institute for Space Research, Sorbonnelaan 2,
NL-3584 CA Utrecht, the Netherlands
\and
Institute of Theoretical Physics and Astrophysics, Kiel University, Leibnizstr. 15, 24118 Kiel, Germany
    }

   \date{...}

% \abstract{}{}{}{}{} 
% 5 {} token are mandatory
 
  \abstract
   {VX~Sgr is a cool, evolved, and luminous red star whose stellar parameters are difficult to determine, which affects its classification.}
   {We aim to spatially resolve the photospheric extent as well as the circumstellar environment.}
   {We used interferometric observations obtained with the MATISSE instrument in the $L$ (3--4\,$\mu$m), $M$ (4.5--5\,$\mu$m), and $N$ (8--13\,$\mu$m) bands. We reconstructed monochromatic images using the MIRA software. We used 3D radiation-hydrodynamics (RHD) simulations carried out with CO$^5$BOLD and a uniform disc model to estimate the apparent diameter and interpret the stellar surface structures. Moreover, we employed the radiative transfer codes {{\sc Optim3D}} and \textsc{Radmc3D} to compute the spectral energy distribution for the $L$, $M,$ and $N$ bands, respectively.}
  % conclusions heading (optional), leave it empty if neompaessary 
   {MATISSE observations unveil, for the first time, the morphology of VX~Sgr across the $L$,  $M$, and $N$ bands. The reconstructed images show a complex morphology with brighter areas whose characteristics depend on the wavelength probed. We measured the angular diameter as a function of the wavelength and showed that the photospheric extent in the $L$ and $M$ bands depends on the opacity through the atmosphere. In addition to this, we also concluded that the observed photospheric inhomogeneities can be interpreted as convection-related surface structures. The comparison in the $N$ band yielded a qualitative agreement between the $N$ -band spectrum and simple dust radiative transfer simulations. However, it is not possible to firmly conclude on the interpretation of the current data because of the difficulty in constraing the model parameters using the limited accuracy of our absolute flux calibration.}
   {MATISSE observations and the derived reconstructed images unveil the appearance of VX~Sgr's stellar surface and circumstellar environment across a very large spectral domain for
the first time.}   
    
   \maketitle
%
%-------------------------------------------------------------------

\section{Introduction}\label{introduction}

VX~Sagittarii (HD~165674, VX~Sgr) is a cool and luminous evolved star. It is a semi-regular variable \citep[732-day period, ][]{1987IBVS.3058....1K} with unusually large magnitudes spanning from a spectral type M5.5 to M9.8 at the visual maximum and minimum phases, respectively \citep{1982MNRAS.198..385L,2006MNRAS.372.1721K}. VX~Sgr has been categorised as an oxygen-rich star \citep{2000A&AS..146..437S} with an effective temperature ranging over three values: 2900\,K \citep{2007A&A...462..711G}, 
3150\,K \citep{2017MNRAS.466.1963L}, and $\sim$3700\,K \citep{2013A&A...554A..76A}. VX~Sgr manifests a strong, asymmetric, and variable mass loss with rates of $\sim$[1$-$6]$\times 10^{-5}$\,$M_\odot$ yr$^{-1}$ \citep{1986MNRAS.220..513C,2010A&A...523A..18D,2011A&A...526A.156M,2017MNRAS.466.1963L,2018AJ....155..212G, 2020A&A...644A.139G}. \\
These mass ejections cause a pronounced extended atmosphere in the $K$ band \citep{2004ApJ...605..436M, 2010A&A...511A..51C} where interferometric observations revealed evidence of departure from circular symmetry, which could be caused by the opacity (H$_2$O and CO) throughout the atmosphere \citep{2010A&A...511A..51C}. In addition to this, its circumstellar environment is detected with SPHERE \citep{2020MNRAS.494.3200S}, where the authors reveal the presence of ejecta confined to non-spherical or clumpy configurations. Moreover, this environment is expected to be characterised by the presence of alumina, metallic iron, and Fe-containing silicates \citep{2020A&A...644A.139G}. Observations in the radio domain, with H$_2$O and SiO masers \citep{2018NatCo...9.2534Y} or OH \citep{1977ApJ...214...60R}, showed larger expansion layers with respect to the typical semi-regular variables.

Radio observations also made it possible to determine precise distances, even if the measures do not entirely overlap among the different works. Using the proper motions of the SiO maser at 43 GHz, 
\cite{2018ApJ...853...42S} reported a distance of 1.10$\pm0.11$ kpc, and  \cite{2007ChJAA...7..531C} found 1.57$\pm0.27$ kpc. Using a 22 GHz maser's emission of H$_2$O, \cite{2018ApJ...859...14X} measured a 
distance of 1.56$^{+0.11}_{-0.10}$ kpc. Other authors estimated the distance with different methods: \cite{1972ApJ...172...75H} placed VX~Sgr in the vicinity of the Sgr OB1 cluster at 1.7 kpc; \cite{1974ApJ...188...75H} calculated a photometric distance of 0.8 kpc; and Hipparcos parallaxes gave a value of 0.262$^{+0.655}_{-0.109}$\,kpc \citep{2007A&A...474..653V}. It should be noted that the latter two measurements could be highly contaminated by convection-related stellar surface structures \citep{2011A&A...528A.120C,2018A&A...617L...1C}. 
As a consequence of these uncertainties, the radius of VX~Sgr ranges between 1000 and 2000 $R_\odot$ \citep{2004ApJ...605..436M,2010A&A...511A..51C,2018ApJ...859...14X} as well as the luminosity between 1.95$\times10^5$ and 5.5$\times10^5$\,$L_\odot$ \citep{2005ApJ...628..973L,2010A&A...511A..51C,2018ApJ...859...14X}, depending on the distance assumed. 

Due to the limitations on precisely determining the stellar parameters, the classification of VX~Sgr is still under debate. On one side, if one assumes the distances obtained in the radio, the star appears to be in agreement to what is expected for standard red supergiant (RSG) stars \citep{2015A&A...575A..50A}; but, on the other hand, \cite{2021A&A...646A..98T} provided new insights into its luminosity, evolutionary stage, and its pulsation period, which seem to point towards some sort of an extreme AGB star. 

In this context, long-baseline infrared interferometric observations play an important role because they allow us to explore and characterise the geometrical extent of the stellar photosphere as well as its circumstellar environment. However, from an observational point of view, RSGs may bear similarities with AGBs \citep[e.g. ][]{2015A&A...575A..50A}, even though they largely differ in terms of mass, effective temperature, surface gravity, and photometric variability \citep{2005ApJ...628..973L,2018A&ARv..26....1H}. \\
In this work we report observations of VX~Sgr obtained with the MATISSE instrument at VLTI and aim to explore the properties of the stellar surface and circumstellar environment in the wavelength bands: $L$,  $M$, and $N$.

\begin{table*} [!h]
\tiny
\begin{center}
 \caption{Summary of all the data obtained for VX~Sgr and its calibrators.}
 \label{log}
 \begin{tabular}{cccccccc}
\hline
  Date & UT\tablefootmark{a}  & Seeing science \tablefootmark{b} & tau0 science \tablefootmark{c} & Calibrators\tablefootmark{d} & Seeing calibrator & tau0 calibrator& AT configuration  \\
          &                                                   & [arcsec]                            & [ms]           &  & [arcsec]                            & [ms]   & \\
\hline
  2019-06-09        & 04:33:19  &  0.6 & 2.8   & $\nu$~Oph&  0.6 & 3.4 & K0G2D0J3/Compact  \\
  2019-06-09        & 05:41:38   &  0.4 & 3.7   &$\beta$~Sgr/$\nu$~Oph  &  1.2/0.7 & 2.1/2.2 & K0G2D0J3/Compact  \\
  2019-06-09        & 07:21:32  &  0.4 & 3.7   & $\nu$~Oph &  1.2 & 2.1 & K0G2D0J3/Compact  \\
  2019-06-09        & 08:58:50  &  1.6 & 2.0   & $\beta$~Sgr &  1.1 & 2.8 & K0G2D0J3/Compact  \\
  \hline
  2019-06-25        & 01:02:23  &  0.6 & 2.8  & $\delta$~Sgr &  0.9 & 0.9 & A0B2D0C1/Medium  \\
  2019-06-25        & 02:41:26  &  1.1 & 0.7  & $\delta$~Sgr &  1.1 & 1.1 & A0B2D0C1/Medium   \\
  2019-06-25        & 03:38:24  &  1.0 & 0.8  & $\delta$~Sgr &  1.0 & 0.7 & A0B2D0C1/Medium   \\
  2019-06-25        & 04:33:20  &  1.0 & 0.8  & $\beta$~Sgr/$\delta$~Sgr &  1.1/1.0 & 0.8/1.0 & A0B2D0C1/Medium   \\
  2019-06-25        & 06:12:47  &  1.1 & 1.0  & $\delta$~Sgr &  1.1 & 0.8 & A0B2D0C1/Medium   \\
  2019-06-25        & 07:04:58  &  1.2 & 0.8  & $\delta$~Sgr &  1.2 & 1.0 & A0B2D0C1/Medium   \\
  2019-06-26        & 00:42:32 &  1.6 & 0.8  & $\delta$~Sgr &  1.3 & 0.8 & A0B2D0C1/Medium   \\
\hline
  2019-08-29        & 00:16:25 &  0.5 & 3.3 & $\beta$~Sgr/$\nu$~Oph/$\alpha$~Ser &  0.9/0.8/1.4 & 1.5/1.7/1.5 & A0G1J2J3/Large  \\
\hline
 \end{tabular}
\end{center}
\tablefoot{\\
\tablefoottext{a}{UT time at the start of the data acquisition.}\\
\tablefoottext{b}{Seeing during the observation for the science target.}\\
\tablefoottext{c}{Coherence time in the visible spectral region of the science target.}\\
\tablefoottext{d}{Calibrator stars used. Some parameters of the calibrators are reported in Table~\ref{calibrators}.}}
\end{table*} 

\begin{table*} 
\begin{center}
 \caption{\emph{\emph{Left:}} Parameters of the calibrators for VX~Sgr extracted from \cite{2019MNRAS.490.3158C}. \emph{\emph{Right:}} Stellar parameters of the synthetic spectra \citep[PHOENIX grid, ][]{2013A&A...553A...6H} used for the calibrators (see text): effective temperature 
($T_\mathrm{eff}$), surface gravity ($\log g$), and metallicity $[Fe/H]$.}
 \label{calibrators}
 \begin{tabular}{ccc|ccc}
\hline
  Name & Spectral type  & LDD\tablefootmark{a} Diameter & $T_\mathrm{eff}$ & $\log g$ & $[Fe/H]$ \\ 
            &                         &      [mas]            &           [K]         &    [cgs] & \\     
\hline
HD 163917 ($\nu$~Oph) & G9IIIa  & 2.789$\pm$0.005 & 5100 & 2.5 & +0.5 \\
HD 188603 ($\beta$~Sgr) & K2.5IIb  & 2.895$\pm$0.013 & 4200 & 1.0 & +0.5 \\
HD 168454 ($\delta$~Sgr) & K2.5IIIa & 5.874$\pm$0.026 & 4600 & 1.5 & +0.5 \\
%HD 192947 ($\alpha^2$~Cap) & G9III & 2.357$\pm$0.011 & 5200 & 2.5 & +0.5 \\
HD 140573 ($\alpha$~Ser) & K2IIIb & 4.770$\pm$0.013 & 4200 & 3.0 & 0.0 \\
\hline
 \end{tabular}
\end{center}
\tablefoot{\tablefoottext{a}{Limb-Darkened Disc (LDD)}}
\end{table*} 

%--------------------------------------------------------------------
\section{Interferometric observations}

\subsection{Short description of the instrument}

Multi AperTure Interferometric and SpectroScopic Experiment \citep[MATISSE; ][]{2021arXiv211015556L} is the second-generation four-telescope interferometric beam combiner of the Very Large Telescope Interferometer (VLTI), which enables milliarcsecond(s) angular resolution imaging in the $L$ (3--4\,$\mu$m), $M$ (4.5--5\,$\mu$m), and $N$ (8--13\,$\mu$m) mid-infrared bands. It provides exquisite spatial resolution: $\lambda/2B = 2.25$\,mas, and $\lambda/B = 7.50$\,mas for a baseline of $B=140$\,m and for wavelengths $\lambda=3\,\mu$m and $\lambda=10\,\mu$m, respectively. Moreover, it can provide spectrally dispersed data at low ($R=\lambda/\delta\lambda=$35, the one used in this work), medium ($R=$500), high ($R=$960), and high+ ($R=$3600) spectral resolutions in $L$ and/or $M$ bands, and low ($R=$35) or high ($R=$300) in $N$ band.

In each of the $L$,  $M,$ and $N$ bands, MATISSE uses a multi-axial all-in-one beam combination, that is the four separated telescope beams are focussed onto the detector by a common lens and produce six dispersed fringe patterns that are mixed into one single focal spot (an airy disc), representing the so-called interferometric channel.
In the $L$ and $M$ bands, the total incoming flux is split between the interferometric channel and the four photometric channels (SiPhot mode). This mode allows us to simultaneously perform the source photometry and thus to properly calibrate the visibilities.
In the $N$ band, the source photometry is obtained after the fringes recording by closing alternatively the MATISSE shutters to feed the four photometric channels. That allows us to improve the sensitivity by sending 100$\%$ of the total incoming flux in the different channels (high sens mode).

\begin{figure}
   \centering
    \begin{tabular}{c}
     \includegraphics[width=1\hsize]{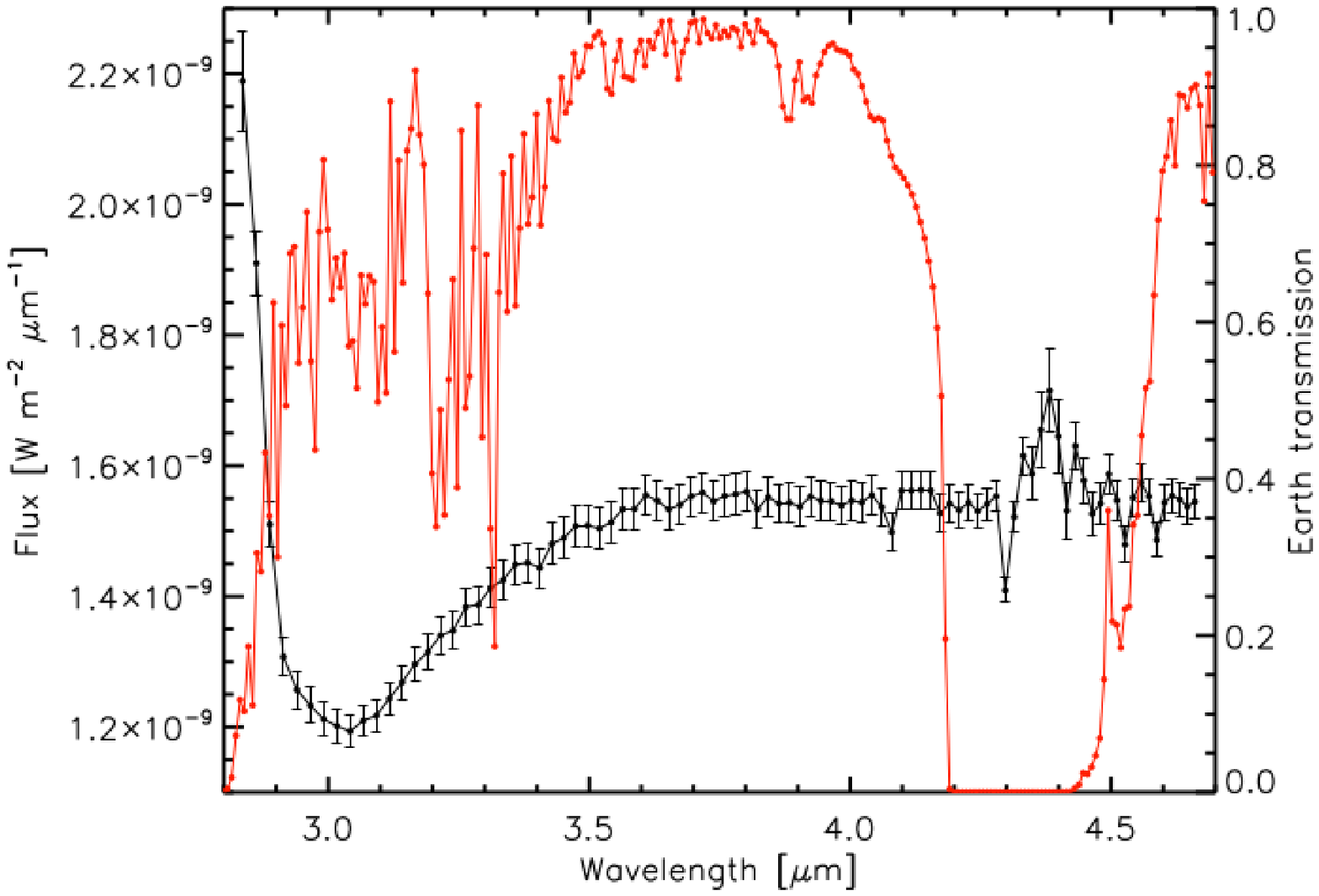} \\
     \includegraphics[width=1\hsize]{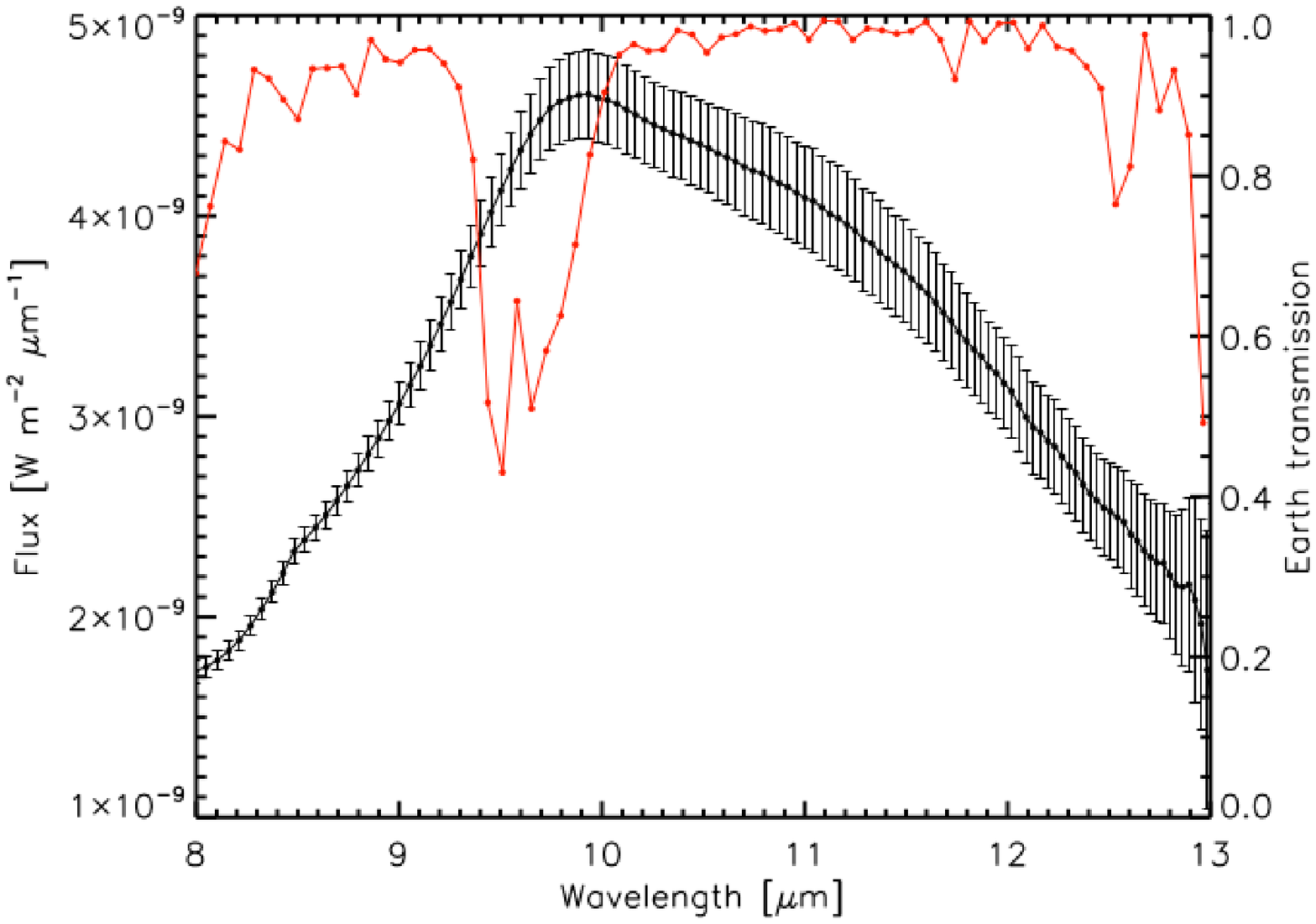} 
      \end{tabular}
   \caption{Black curve is a representative example of calibrated MATISSE flux of VX~Sgr (see Sect.~\ref{datareductiontrue}) in $L$ and $M$ bands (\emph{top}) and $N$ band (\emph{low}) overplotted to the telluric spectrum at the observing site (red) computed with the ESO Skycalc web application and sampled to the same resolving power of the MATISSE observations. The night shown in this plot is 25 June 2019.}
        \label{example_water}
\end{figure}  

MATISSE operates in the mid-infrared, where thermal background emission is significant. As a consequence, two additional modulation steps have been added to calibrate the data: the first one is chopping modulation (observing the target and an empty area of the sky alternatively). It is a mandatory step to extract all source photometries in N band while allowing reliable photometry measurements for faint targets in the $L$ and $M$ bands. The second modulation is an optical path difference (OPD) modulation, aimed at removing the low-frequency peak contamination, containing the contribution from the thermal background, to the fringe peaks (in the Fourier transform of the interferograms). The different modes of the instrument (telescope chopping, detector integrations, OPD modulation) are synchronised all together to ensure minimum overheads when observing.
Finally, a beam modulation (or beam commutation) is applied to the observations by swapping two-by-two beams in a four-step sequence. This last feature is aimed at removing detection artifacts in the phases measured by the instrument (differential phases and closure phases). These artifacts are instrument phase residuals located in the optical train between the BCD and the detector \citep{2021arXiv211015556L}. The BCD (Beam Commuting Device) is a module used to calibrate closure and differential phases.

\subsection{Data acquisition and reduction}\label{datareductiontrue}

\begin{figure}
   \centering
    \begin{tabular}{c}
     \includegraphics[width=0.97\hsize]{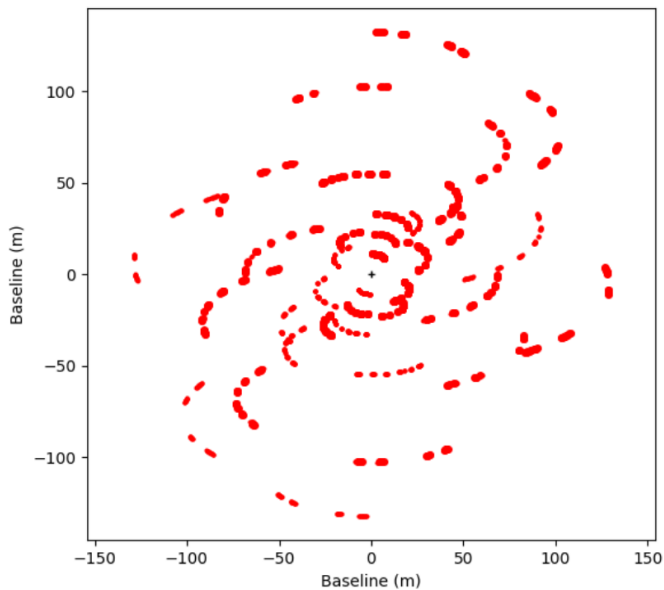} \\
      \end{tabular}
      \caption{UV coverage. The spectral coverage contains wavelengths between 2.8 and 13.0 $\mu$m. The resolution of the interferometer corresponds to a beam of a $132\times132$ meters telescope, $3.1\times3.1$\,mas and $7.8\times7.8$,mas at 4.0 and 10.0\,$\mu$m, respectively.}
        \label{uvplane}
\end{figure}

We observed VX~Sgr at its visible photometric minimum over three nights (Table~\ref{log}) with the MATISSE instrument with the aim of reconstructing images across the different bands. The data collection spanned 2.5 months: the compact and medium configurations within 15 days in June, while the large configuration could only be observed late at the end of August 2019. The calibrators used are reported in Table~\ref{calibrators}, and the UV coverage is displayed in Fig.~\ref{uvplane}. We used the low spectral resolution mode ($R=$35) in the $L$, $M$, and $N$ bands. 

For the data reduction, we applied the MATISSE data reduction software (\texttt{matisse-drs}, \citealt{2016SPIE.9907E..23M}). It adopts the classical Fourier transforms scheme \citep{Perrin2003}, with photometric calibration in the $L$ and $N$ bands \citep{Foresto1997}. Chopping correction (subtracting sky frames from target frames) and OPD demodulation are applied before summing spectral densities\footnote{Squared modulus of the Fourier transform of the interferograms.}, bispectra\footnote{The product of the complex coherent fluxes (i.e. the Fourier transform of the interferograms) over a baseline triplet. The argument of such quantity is the closure phase.}, and interspectra \footnote{The product of the complex coherent flux with the complex coherent flux measured at a specific wavelength (or a mean of the complex coherent flux over the spectral bandwidth). The argument of such quantity is the differential phase.}, leading to estimates of the spectrointerferometric observables; spectrally dispersed in squared visibility, closure phase, and differential phase, similarly to what was previously done with the AMBER instrument \citep{2007A&A...464....1P,2007A&A...464...29T}. A MATISSE observation sequence is composed of twelve one-minute-long exposures with simultaneous interferometric and photometric data in the $L$ $and M$ bands and four interferometric exposures followed by eight photometric exposures in the $N$ band. The first four exposures are taken without chopping while the following eight exposures are chopped. Each exposure is taken in one of the four configurations of the BCD. Overall, this leads to four reduced OIFITS2 \citep{2017A&A...597A...8D} files per pointed target (one for each BCD configuration) containing, per exposure, six dispersed squared visibilities, differential visibilities phases, and three independent dispersed closure phases (out of four in total). The interferometric calibration is then applied to the data: the targets used as calibrators (CAL) are corrected from their expected observables given the predicted value of their angular diameters, and each science target (SCI) is corrected from the transfer function thus estimated. The calibrated OIFITS2 files of the four BCD configurations are merged to produce a single calibrated OIFITS2 file for each CAL-SCI pair.  

\begin{figure*}
   \centering
    \begin{tabular}{cc}
     \includegraphics[width=0.5\hsize]{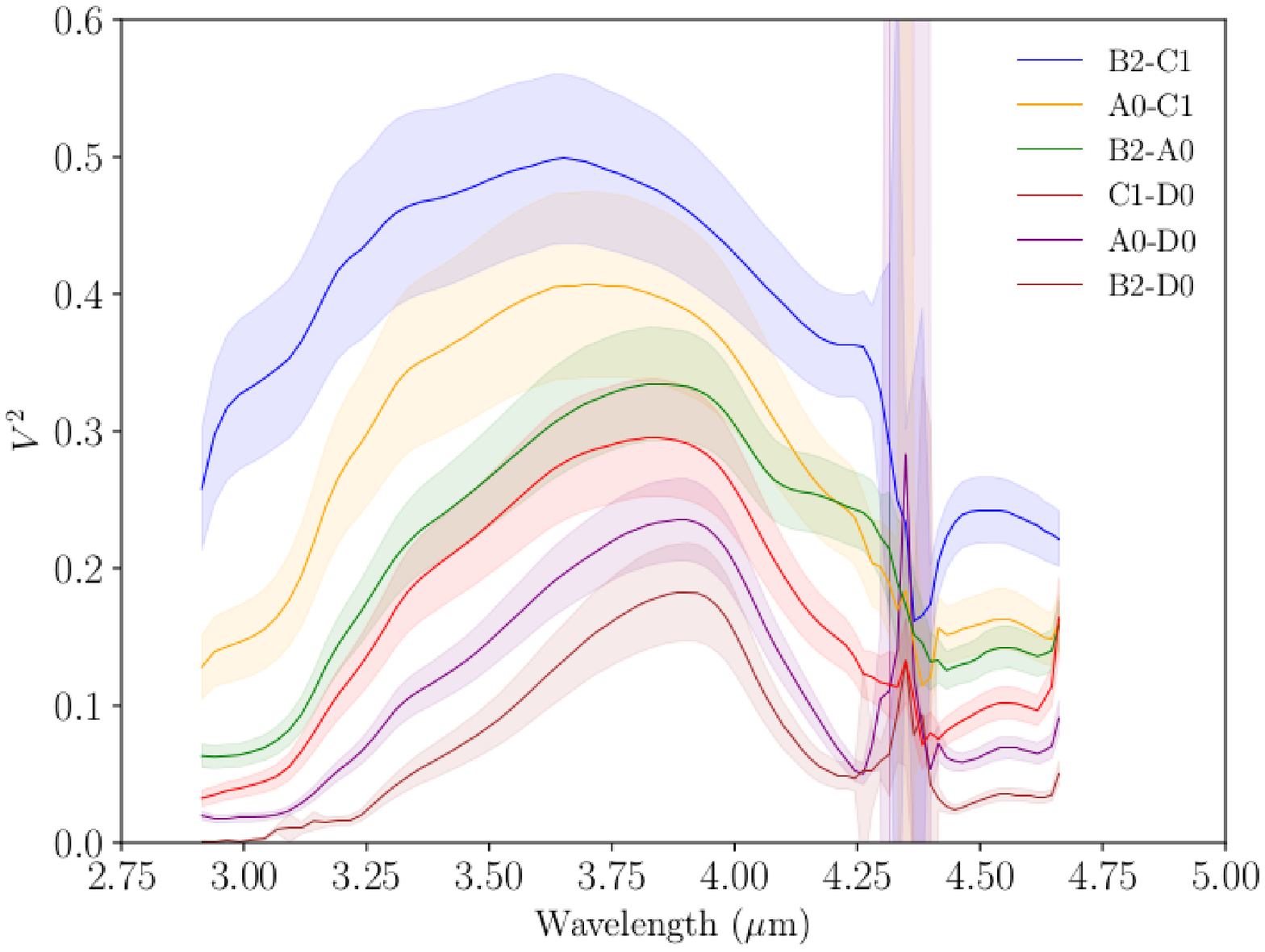} 
     \includegraphics[width=0.5\hsize]{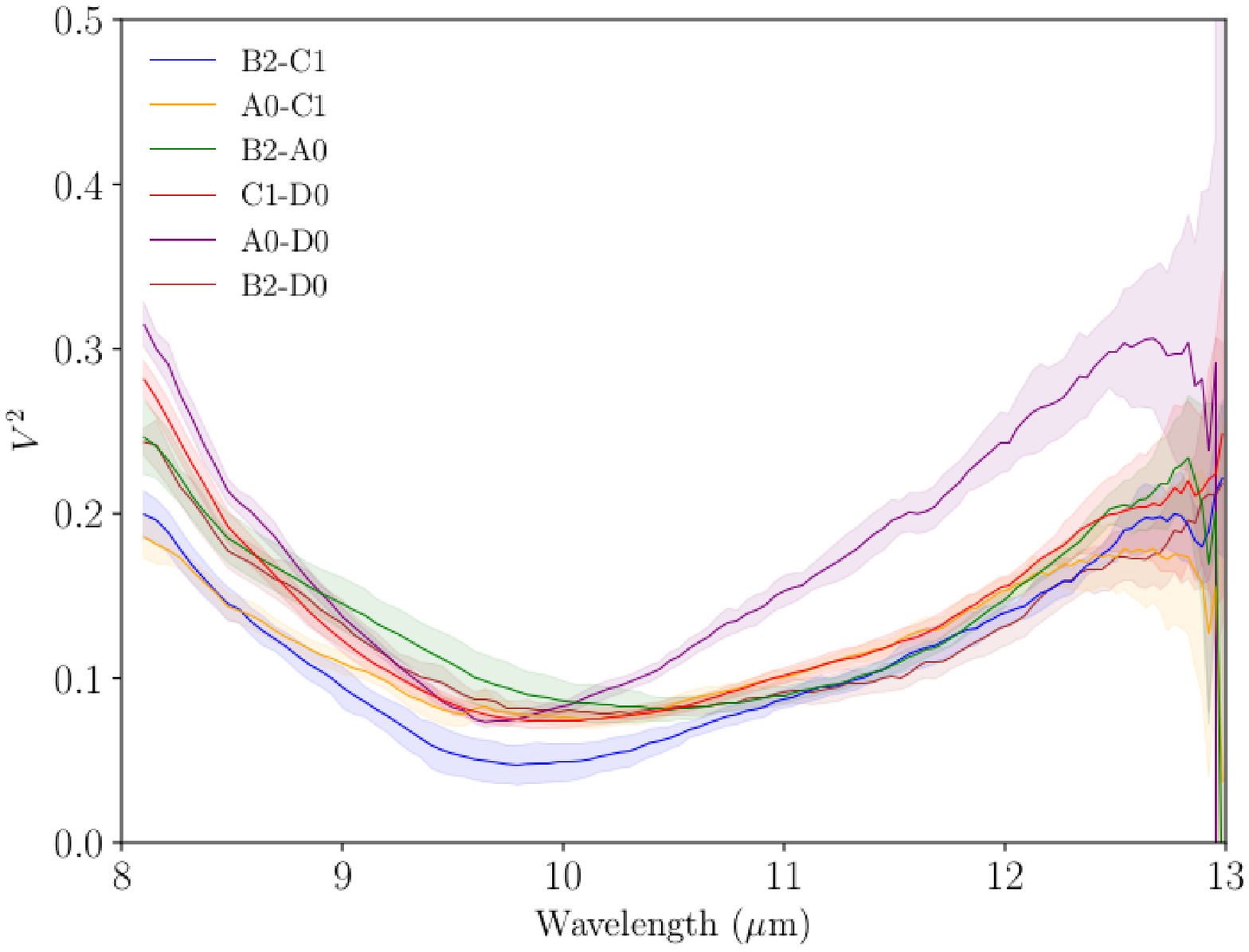} \\
     \includegraphics[width=0.5\hsize]{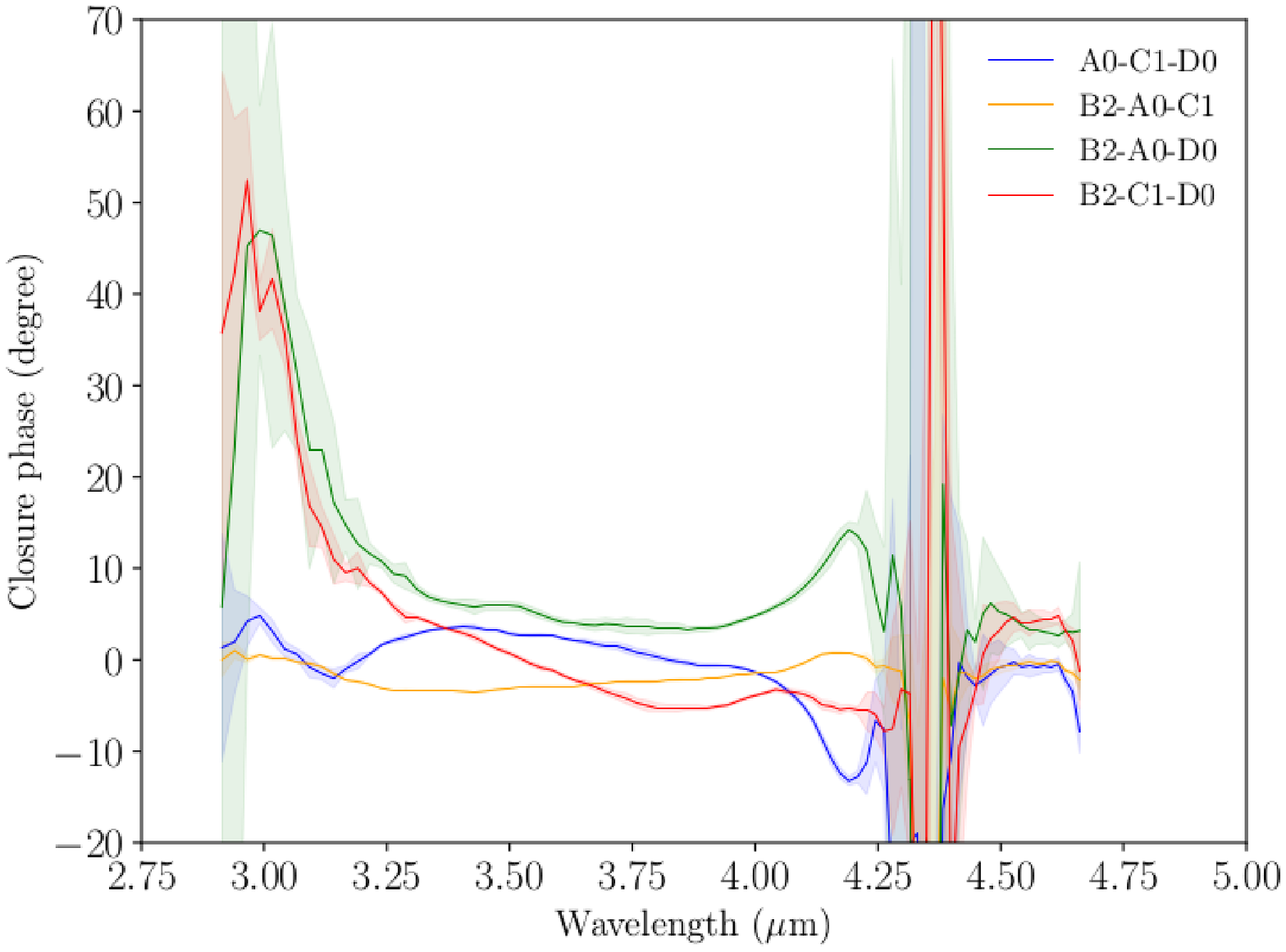} 
     \includegraphics[width=0.5\hsize]{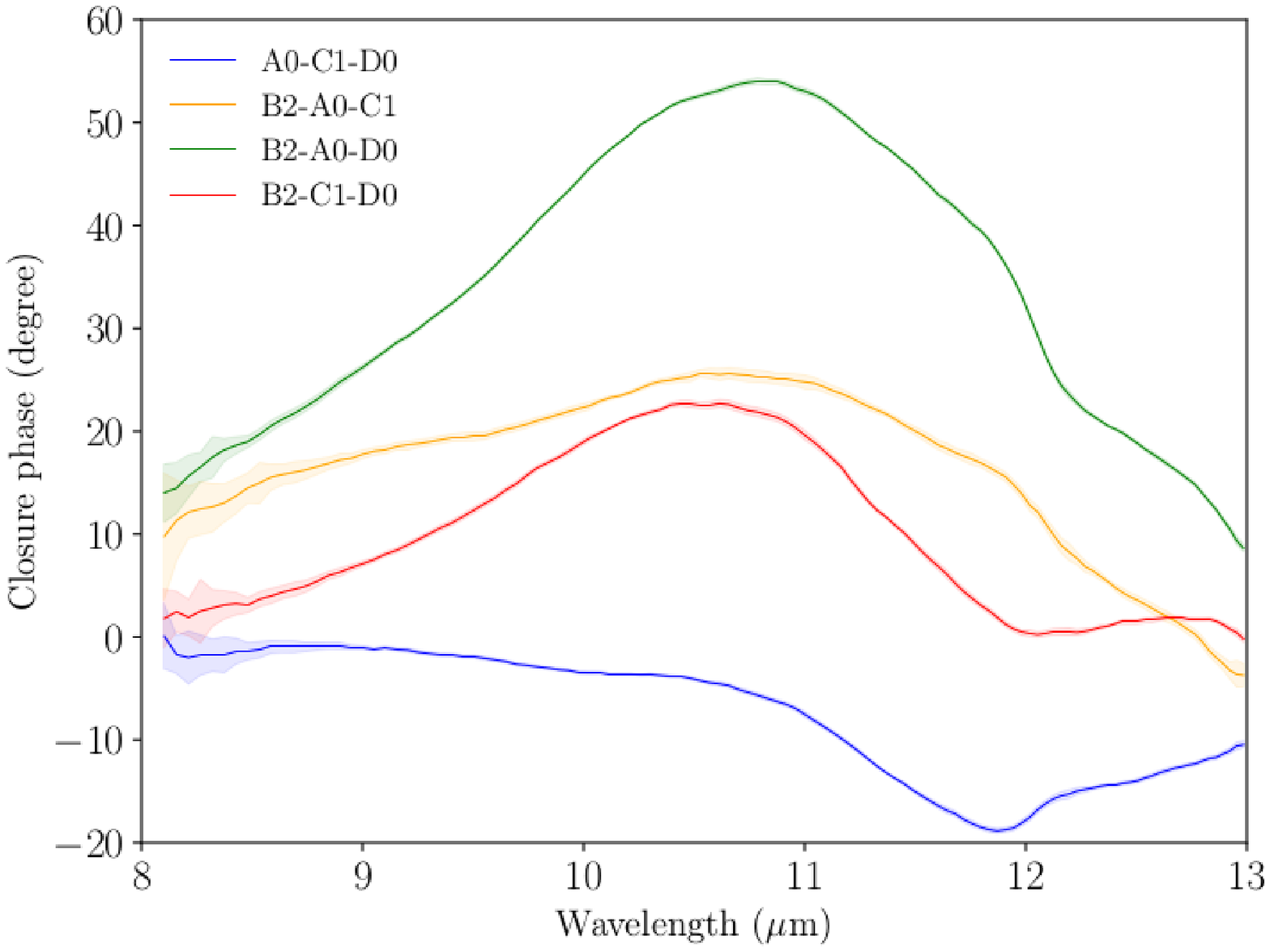} \\
      \end{tabular}
      \caption{Representative example of visibility amplitudes (\emph{top panels}) and closure phases (\emph{bottom panels}) for the $L$ and $M$ bands (\emph{left column}) and the $N$ band (\emph{right column}). The different colours indicate different baselines for visibilities and triplets for closures phases. The spectral region between 4.27 and 4.44 $\mu$m suffers from noise due to the the telluric lines in this range. The night shown in this plot is 25 June 2019, as in Fig.~\ref{example_water}.}
        \label{example_obs}
\end{figure*}

Figure~\ref{example_obs} shows a representative example of the $L$,  $M,$ and $N$ bands' calibrated squared visibilities and closure phases taken with the small AT configuration (A0-B2-D0-C1) on 15 June 2019. The data taken at low spatial frequencies are of good quality, except for the ranges of the atmospheric absorption, and they already show a spatially resolved emission in all the bands, with strong brightness asymmetries in the $N$ band indicated by the significant non-null closure phases. 

Figure~\ref{example_water} displays the calibrated MATISSE spectrum of VX~Sgr for one particular night (25 June 2019) overplotted on the telluric spectrum at the observing site. For each observation block of VX Sgr, the absolute flux calibration was performed by dividing the raw spectrum of VX Sgr by the raw spectrum of the associated calibrator and then multiplying this ratio by the calibrator synthetic spectrum. For the night shown in the figure, the synthetic spectrum of the calibrator $\delta$~Sgr (Table~\ref{calibrators}) comes from the PHOENIX grid \citep[ACES-AGSS-COND,][]{2013A&A...553A...6H}. According to Fig.~\ref{example_water}, the data quality is good, while the telluric line dominates in specific spectral regions: lower than $\sim$3.3 $\mu$m, between 4.0 and 4.7 $\mu$m, and between 9.3 and 9.9 $\mu$m. In the following, we include all spectral regions in our work, but we highlight the telluric spectrum at the observing site.

%-------------------------------------- Two column figure   

\section{The observed stellar surface of VX~Sgr}
 
We performed image reconstruction for the observed data using the MIRA software package developed by \cite{2008SPIE.7013E..1IT}. This package is written in the scientific language yorick\footnote{\url{https://software.llnl.gov/yorick-doc/}}. It uses interferometric data in the form of OIFITS2 files \citep{2005PASP..117.1255P}. The reconstructed image is compared to the visibility and closure phase data by means of Fourier transforms (FT): a FT is first computed on the image, and visibilities are computed as the FT amplitude at the observation spatial frequencies ($f_{ij}=\frac{B_{ij}}{\lambda}$), $B_{ij}$ being the projected baseline between telescopes $i$ and $j$, and $\lambda$ the wavelength of observation. On the other hand, phases are computed as the FT phase at $f_{ij}$. The closure phases are computed as the sum of the three phases at the three observation spatial frequencies ($f_{ij}, f_{jk}, f_{ki}$) corresponding to a telescopes triangle. The pixels values are modulated in intensity to minimise the so-called
"cost function", which is the sum of a regularisation term plus
data-related terms ($\chi^2$). The data terms enforce the agreement of the model
image with the measured data (visibilities and closure phases).
The regularisation term is a distance minimisation between the
reconstructed image and an expected image, called the prior. The data term and the regularisation terms must be scaled together (thanks to the use of the "hyperparameter" ($\mu$\footnote{The hyperparameter is used to adjust the weight of the constraints set by the measurements relatively to the ones set by the priors}) to keep a similar magnitude in the characterise process. 

\begin{figure*}
   \centering
    \begin{tabular}{cc}
     \includegraphics[width=0.5\hsize]{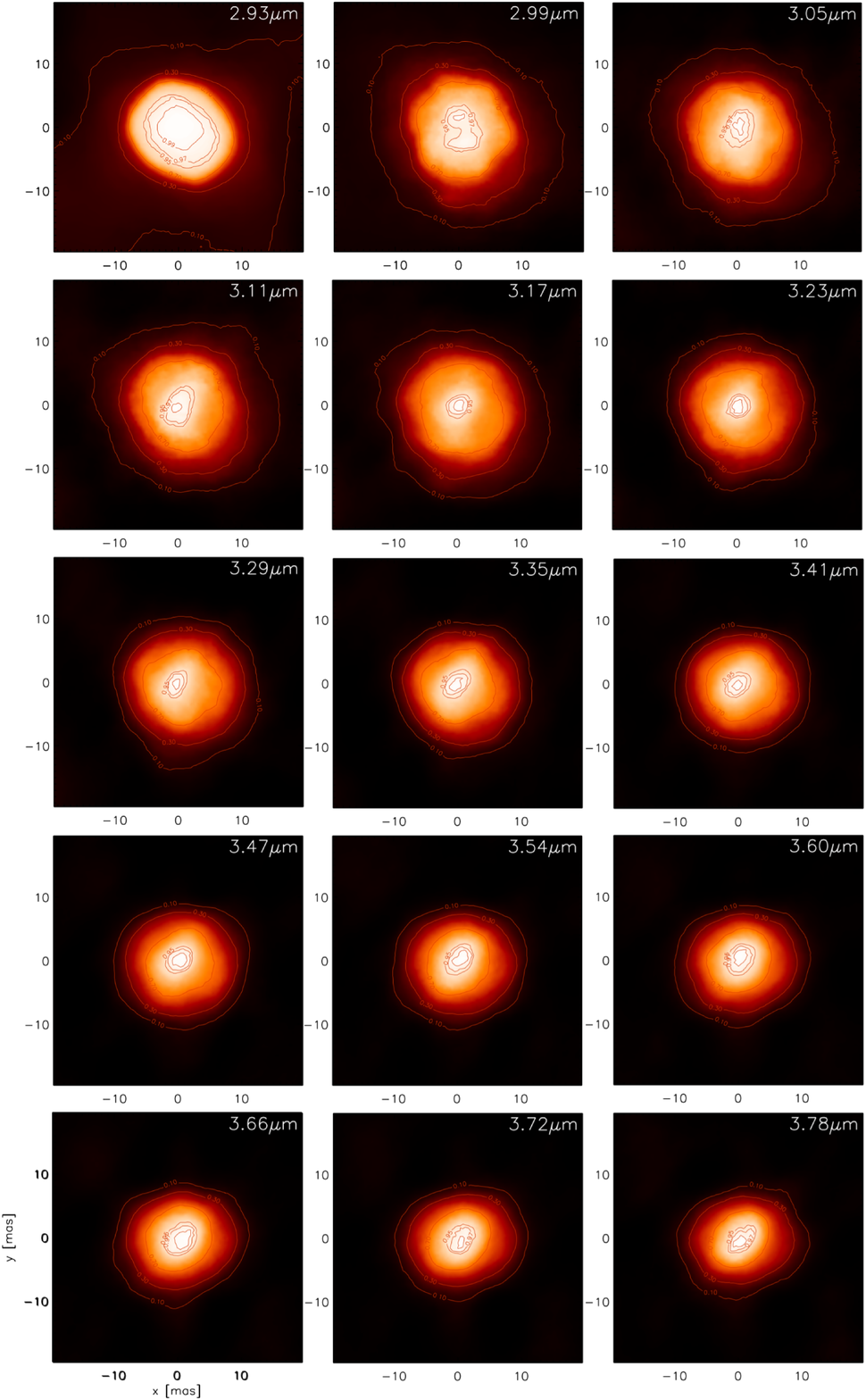}
     \includegraphics[width=0.5\hsize]{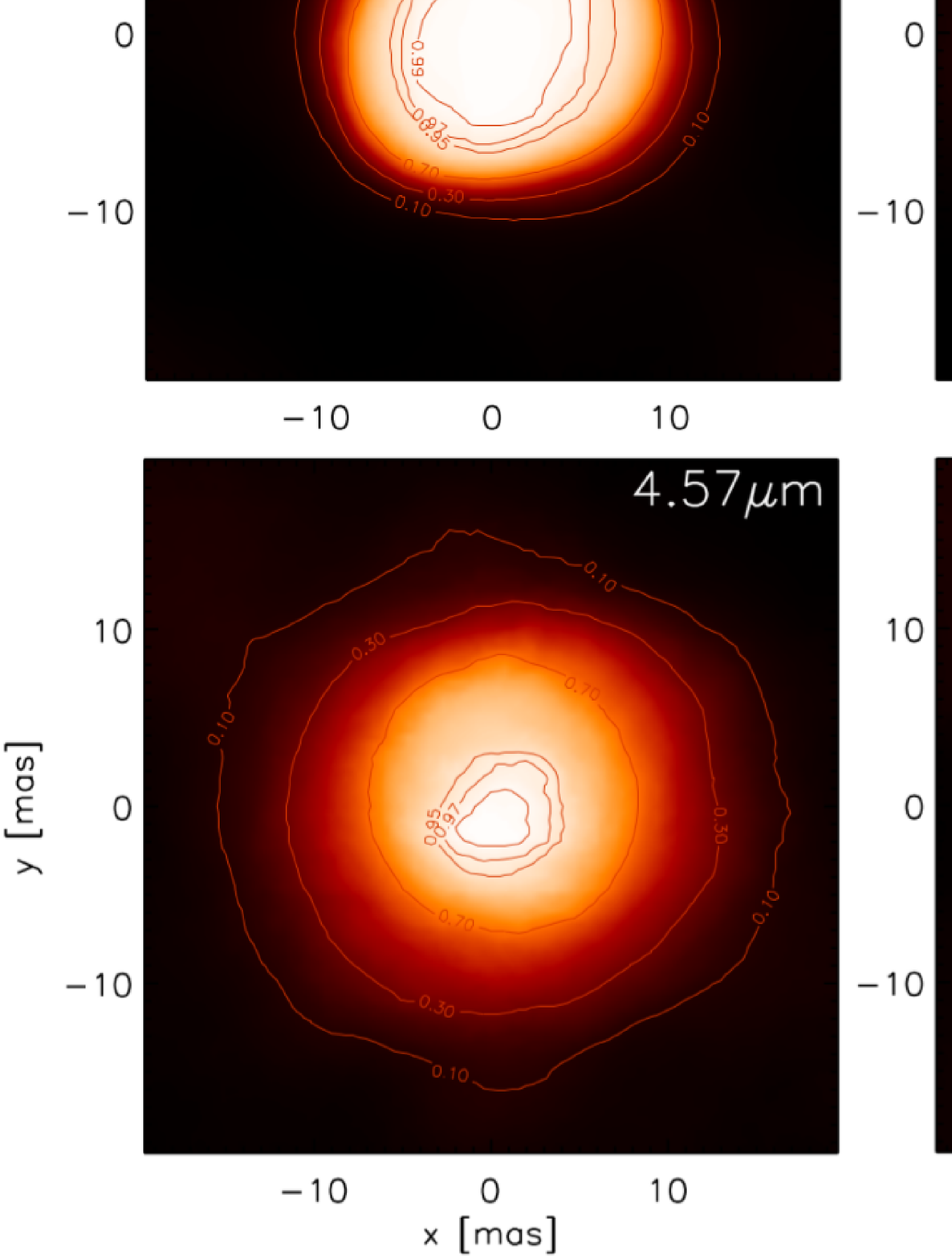}
     \end{tabular}
      \caption{Reconstructed images in $L$ and $M$ bands. They have $256\times256$ pixels with 0.39\,mas pixel$^{-1}$ and are convolved with the interferometric beam (Fig.~\ref{uvplane}). The peak intensity is normalised between [0, 1], and the corresponding noise maps are shown in Fig.~\ref{RMSimage1}. The highest noise value is $\sim1.5\times10^{-4}$. The contour lines correspond to [0.10, 0.30, 0.70, 0.95, 0.97, 0.99]. East is to the left, and north is up. As reported in Section~\ref{datareductiontrue}, the data quality of the images at wavelengths lower than $\sim$3.3 $\mu$m and between 4.0 and 4.7 $\mu$m is compromised by the telluric spectrum at the observing site.}
        \label{image1}
\end{figure*}

\begin{figure}
   \centering
    \begin{tabular}{cc}
      \includegraphics[width=0.97\hsize]{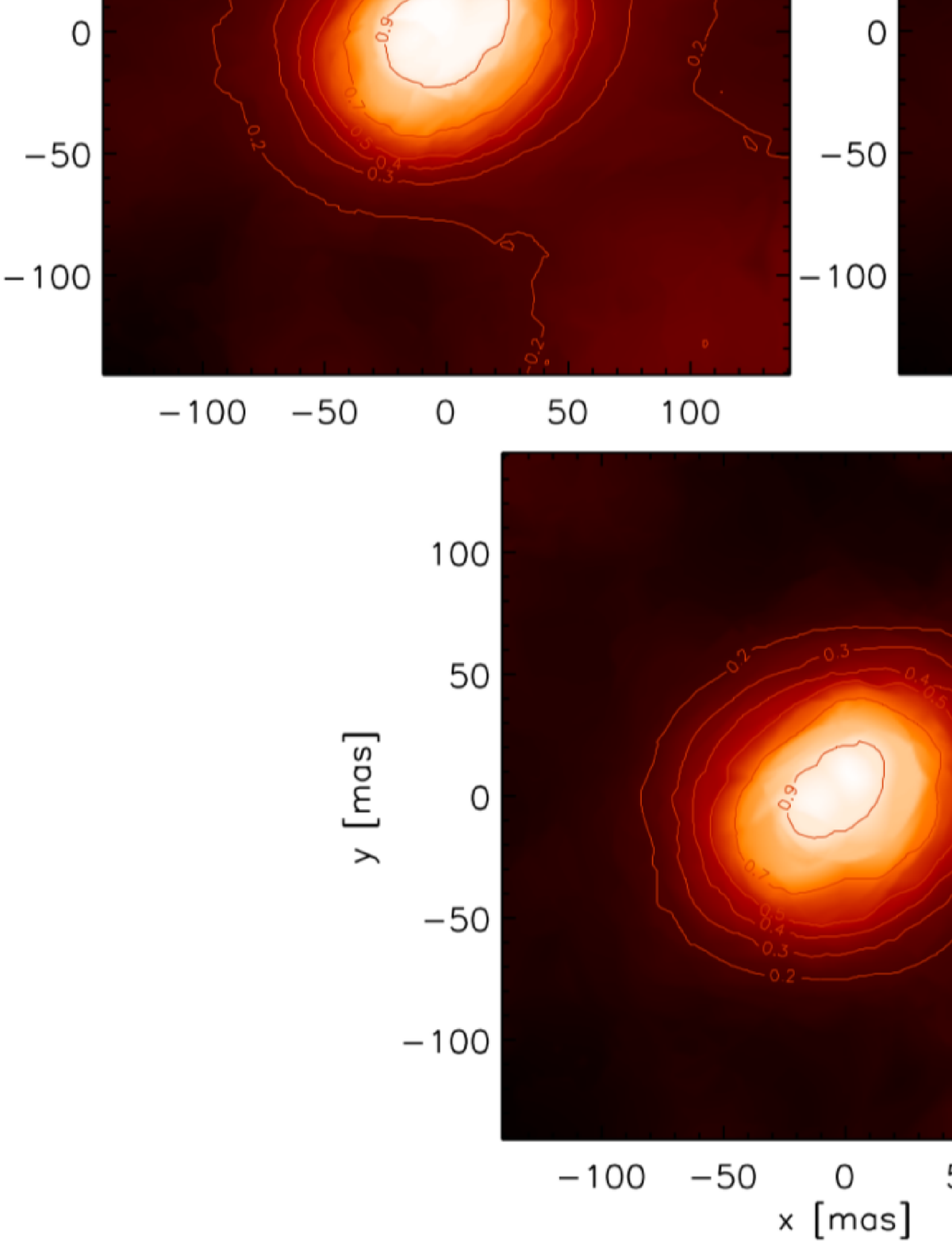}
     \end{tabular}
      \caption{Reconstructed images in the $N$ band. They have $512\times512$ pixels with 0.59\,mas pixel$^{-1}$ and are convolved with the interferometric beam (Fig.~\ref{uvplane}). The peak intensity is normalised between [0, 1] and the corresponding noise maps are shown in Fig.~\ref{RMSimage2}: the highest noise value is $\sim3\times10^{-5}$. The contour lines correspond to [0.20, 0.30, 0.40, 0.50, 0.70, 0.95]. East is to the left, north is up. As reported in Sect.~\ref{datareductiontrue}, the data quality of the images at wavelengths between 9.3 and 9.9 $\mu$m, is compromised by the the telluric spectrum at the observing site.}
             \label{image2}
\end{figure}

\begin{figure*}
   \centering
    \begin{tabular}{cc}
         \includegraphics[width=1.\hsize]{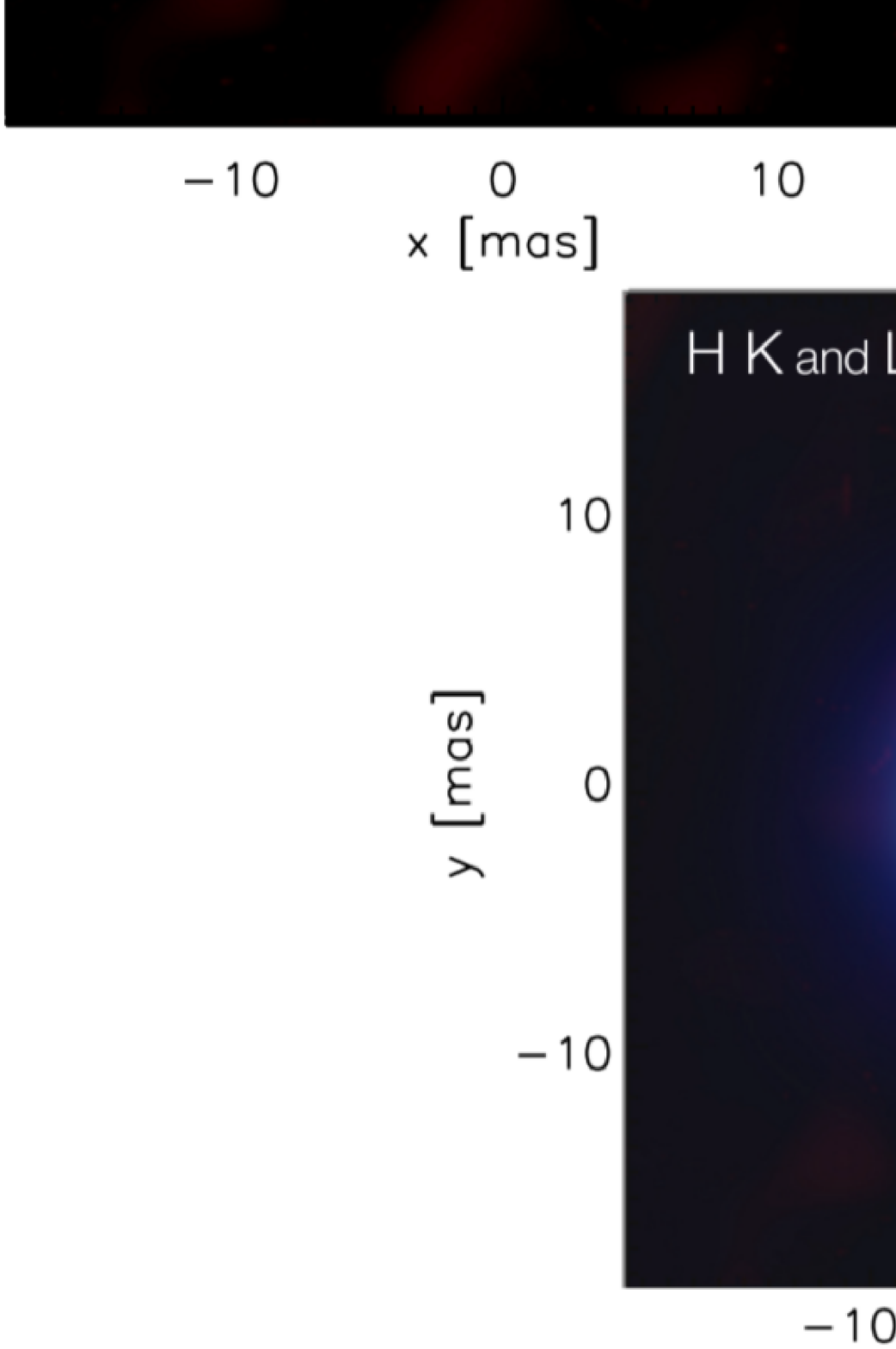}
     \end{tabular}
      \caption{\emph{Top row:} colour composite images in the $H$ and $K$ bands \citep[observation taken in 2008 with the AMBER instrument, with 21 spectral channels combined, ][]{2010A&A...511A..51C}, $L$ and $M$ (31 spectral channels from Fig.~\ref{image1} combined), and $N$ band (17 spectral channels from Fig.~\ref{image2} combined). \emph{Bottom row, left panel:} colour composite images for the $H-K$ (pink scale) and $L-M$ (blue scale) bands. \emph{Bottom row, right panel:} All the bands, including $N$ (green scale), are combined. The photospheric structures (H to M bands) are located in the centre of the N band maps.}
        \label{composite}
\end{figure*}

We reconstructed images using the observed visibilities and closure phases at each spectral channel individually and assuming a resolution of 0.39 and 0.59\,mas pixel$^{-1}$ and a field of view (FOV) of $256\times256$  pixels and $512\times512$\,pixels in the $L$,  $M,$ and $N$ bands, respectively. These represent (i) a total variation regularisation, (ii) a circular Gaussian prior image with a full width at half maximum (FWHM) equal to 0.5$\times$FOV, (iii) a hyperparameter value of $\mu=5\times10^4$. For each observation block, the absolute flux calibration was performed by dividing the raw spectrum of the science target by the raw spectrum of the associated calibrator. Fig.~\ref{image1} displays the reconstructed images in the $L$ and $M$ bands, while Fig.~\ref{image2} shows the same in the $N$ band. Fig.~\ref{composite} shows colour composite images across different wavelength ranges and illustrates the much larger structures in the N band compared to shorter wavelengths. The corresponding intensity RMS (i.e. noise maps) for all the images is reported in Figs.~\ref{RMSimage1} and ~\ref{RMSimage2}.

In the $L$ and $M$ bands, the images show a complex morphology with a central brighter area, whose position depends on the wavelength probed. The photocentre deviates from the geometrical centre, as already pointed out for another evolved star using the same reconstruction method \citep{2020A&A...640A..23C}. In addition to this, the stellar apparent diameter is larger at shorter wavelengths, smaller between $\sim$3.5 and $\sim$4.0\,$\mu$m, before getting larger again for longer wavelengths. In the $N$ band, at $\lambda\sim8.40$\,$\mu$m where the brightness is about $10\%$ of the central region. Then the contrast increases with wavelengths to $\sim30\%$ at $\sim$11\,$\mu$m and reaches a maximum of almost $40\%$ at $\sim$13\,$\mu$m.

\section{Interpreting the observations in the $L$ and $M$ bands}

In this section, we present the simulations and methodology used to make a comparison to the observations.

\subsection{3D simulations of AGB and RSG stars}

We used the radiation-hydrodynamics (RHD) simulations of stellar surface convection computed with the CO$^5$BOLD \citep{2012JCoPh.231..919F} code. The code solves the coupled non-linear equations of compressible hydrodynamics and non-local radiative energy transfer in the presence of a fixed external spherically symmetric gravitational field on a 3D cartesian grid. No artificially-induced pulsations are added to the simulations (e.g. by a piston) but they are self-excited. Molecular opacities are taken into account, but radiation transport is treated in a grey approximation, ignoring radiation pressure and dust opacities. Dynamical pressure allows the density to drop much more slowly as a function of the distance than expected for a hydrostatic atmosphere \citep[for more details see ][]{2017A&A...600A.137F}.

Since the uncertainty on the stellar properties of VX~Sgr is large (Sect.~\ref{introduction}), we used two different simulations: (i) for the AGB stellar type, we used the one from \cite{2017A&A...600A.137F} with the highest $T_\mathrm{eff}$ and $L_\star$, and (ii) for the RSG we used the one presented in detail in \cite{2019A&A...632A..28K}. The averaged stellar parameters of the simulations used in this work are reported in Table~\ref{simus}. The temporal variability, used to increase the number of possible matching images, is ensured by the 100 snapshots (5.75 years, with temporal timestep of $\sim$20 days) for the AGB simulation and 180 (10.35 years, with temporal timestep of $\sim$20 days) for the RSG one. These snapshots are extracted from the relaxed simulated time sequence reported in Table~\ref{simus}.

Afterwards, we employed {{\sc Optim3D}} \citep{2009A&A...506.1351C} to compute synthetic intensity maps in the same observed spectral channels as MATISSE and using as input the RHD simulations snapshots. This code takes into account the Doppler shifts caused by the
convective motions. The radiative transfer is computed in detail using pre-tabulated extinction coefficients per unit mass, as for the hydrostatic model atmosphere code MARCS (\citealp{2008A&A...486..951G}), as functions of temperature, density, and wavelength for the solar composition \citep{2009ARA&A..47..481A}. The
temperature and density distributions are optimised to cover the values
encountered in the outer layers of the RHD simulations. 

The MATISSE observations cover the $L$,  $M$, and $N$ bands. While the $L$ and $M$ bands are mostly characterised by molecular features, the N-band spectrum may display several dust features \citep[e.g. ][]{2017A&A...600A.136P}, which in turn affect the general shape of the star and its circumstellar environment. On the theoretical side, there are no source terms or dedicated opacities for dust in RHD simulations \citep{2017A&A...600A.137F}. In the end, we used the RSG and AGB models to compute synthetic images in the $L$ and $M$ bands, and, only for a testing purpose, the AGB model for the $N$ band.

\begin{table*}
\footnotesize 
\begin{center}
 \caption{Parameters of the RHD simulations used in this work.}
 \label{simus}
 \begin{tabular}{l|ccccccccc}
\hline
Simulation & Stellar &  $M_\star$ & $L_\star$ & $R_\star$ & $T_\mathrm{eff}$ & $\log g$ & $t_\mathrm{avg}$ & Grid & x$_\mathrm{box}$\\
& type& $[M_\sun]$  & $[L_\sun]$  & $[R_\sun]$ & [K] & [cgs] & [yr] & [grid points] & $[R_\sun]$ \\
\hline
st29gm06n001\tablefootmark{a} & AGB & 1 & 6956.3$\pm$547.4 & 350.43$\pm$11.3 & 2814.7$\pm$68.9 & $-$0.65$\pm$0.03 & 25.35 & 281$^3$ & 1381 \\
st35gm04n38\tablefootmark{b} & RSG & 5 & 41517.3$\pm$1074.4 & 582.03$\pm$4.7 &  3414.2$\pm$16.8 & $-$0.40$\pm$ 0.01 & 11.46 & 401$^3$ & 1631 \\
\hline
 \end{tabular}
 \end{center}
\tablefoot{The first column shows the simulation name, and the following five columns show the stellar parameters such as the mass ($M_\star$), the average luminosity ($L_\star$), radius ($R_\star$), effective temperature 
($T_\mathrm{eff}$), and surface gravity ($\log g$). The different quantities are averaged over spherical shells and epochs (7th column, $t_\mathrm{avg}$). Errors are one standard-deviation fluctuations with respect to the time average \citep[see][]{2011A&A...535A..22C,2009A&A...506.1351C}. The solar metallicity is assumed. \\
\tablefoottext{a}{Simulation from the AGB grid of \cite{2017A&A...600A.137F}.}\\
\tablefoottext{b}{Simulation of the RSG star presented in detail in \cite{2019A&A...632A..28K}.}
}
\end{table*}

\subsection{Apparent diameter across wavelengths}\label{sect_radius}

\begin{figure}
   \centering
    \begin{tabular}{c}
     \includegraphics[width=0.95\hsize]{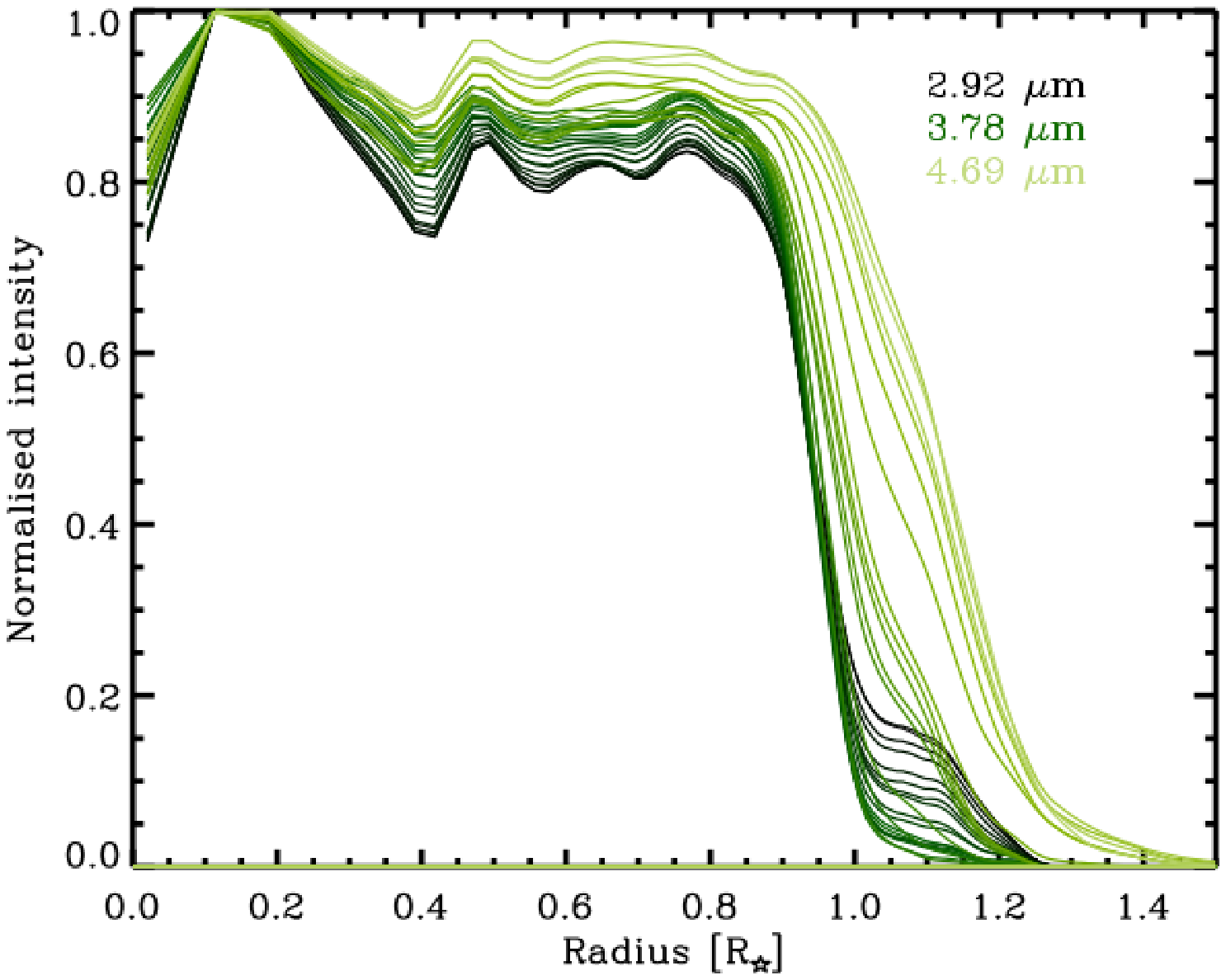}\\
     \includegraphics[width=0.95\hsize]{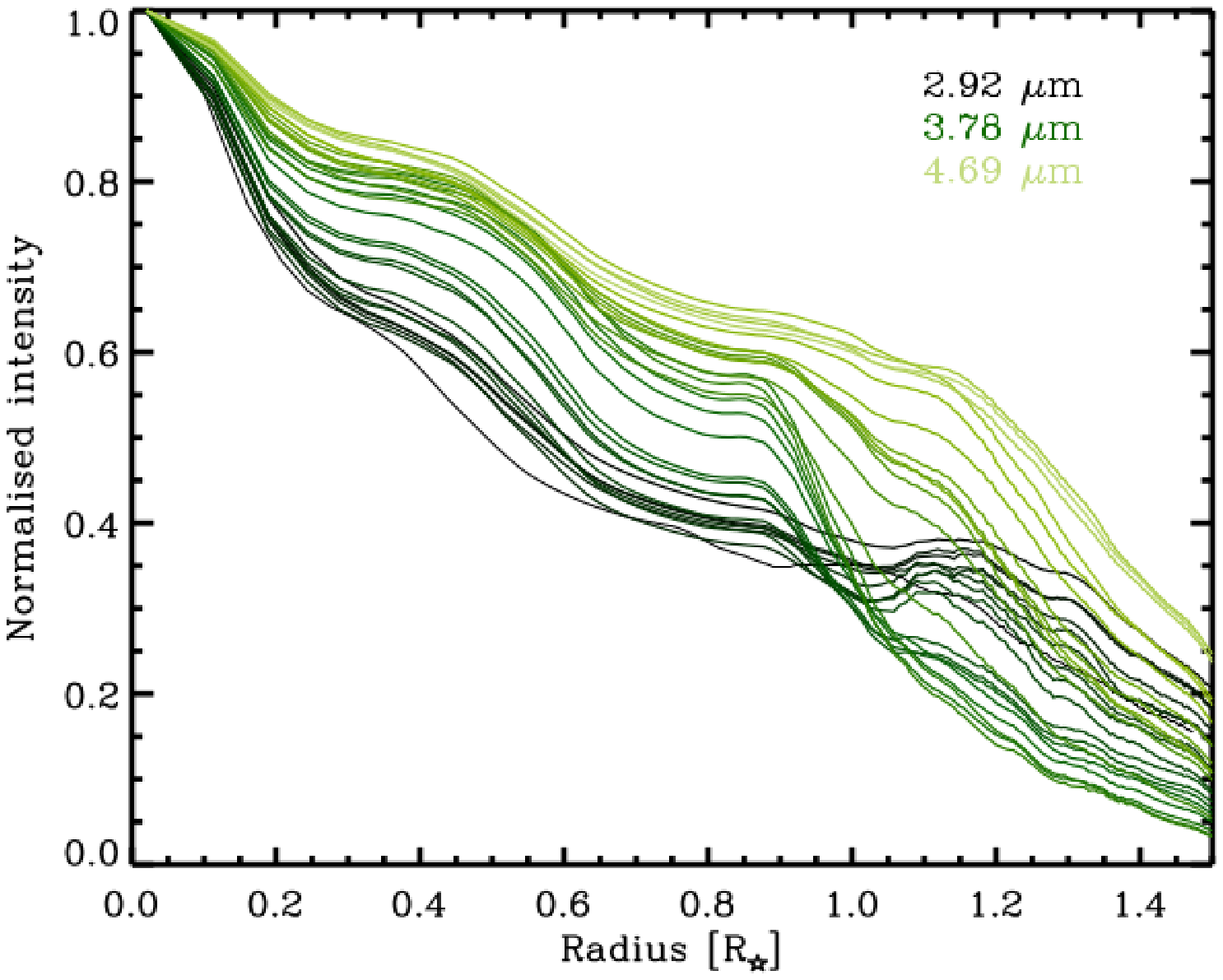}
     \end{tabular}
      \caption{Intensity profiles (normalised to their maximum) as a function of the wavelength in $L$ and $M$ bands for a particular snapshot of the RSG (\emph{top panel}) and the AGB simulations (\emph{bottom panel}) of Table~\ref{simus}. As a reference, few wavelengths are indicated in top right corner.}
        \label{stellar_intensity_profiles}
\end{figure}

We measured the stellar diameters using uniform disk (UD) model and the 3D RHD simulations of Table~\ref{simus}. As a first step, we computed the azimuthally average intensity profiles for all the synthetic maps (i.e. snapshots) of the simulations and for each observed spectral channels. They were constructed using rings regularly spaced in $\mu$ (where $\mu = \cos(\theta)$, with $\theta$ being the angle between the line of sight and the radial direction). Afterwards, we calculated the temporal averages. \\
Figure~\ref{stellar_intensity_profiles} shows examples of snapshots for the RSG and AGB simulations across the $L$ and $M$ bands. These profiles are an effective proxy to investigate the overall shape and extension of their atmospheres. The AGB simulation displays flatter and more extended atmospheres as already shown in \cite{2016A&A...587A..12W}. Moreover, these profiles are also useful to point out the wavelength dependence of the synthetic images: shorter and medium wavelengths (dark and intermediate green) returns smaller radii while longer wavelengths (light green) larger sizes. This aspect is noticeable in the reconstructed images of Fig.~\ref{image1}.

Eventually, we used the temporally averaged (over the simulation time, Table~\ref{simus}) intensity profiles to obtain the synthetic visibility amplitudes using the Hankel transform and fitted the visibility data separately for each observed spectral channel \citep[this procedure has already been used, e.g. in ][]{2020A&A...640A..23C}. The wavelength-dependent apparent diameters of VX~Sgr are reported in Table~\ref{angulardiameter}. As a comparison, \cite{2010A&A...511A..51C} measured a photospheric diameter (corresponding to the reference radius at 1.04\,$\mu$m) of $\Theta=$8.82$\pm$0.50\,mas, which is at least two times smaller than the smallest diameter measured in the $L$ band ($\Theta_\mathrm{AGB}=$18.27$\pm$0.14\,mas at 3.29\,$\mu$m). However, in \cite{2010A&A...511A..51C} the size of VX~Sgr was largely increasing towards the K band up to 2.50\,$\mu$m, where the apparent diameter was close to $\sim$20\,mas. This is very similar to what is found in this work for several spectral channels. In the $K$ band, the authors explained that this increase in size was due to the presence of H$_2$O and CO molecules as dominant absorber in the outer layers.

\begin{table}
\scriptsize 
\begin{center}
 \caption{Apparent diameters (in mas) in the $L$,  $M$, and $N$ bands obtained from a uniform disc ($\Theta_\mathrm{UD}$) model fitted to the MATISSE squared visibilities and fitting the synthetic RSG ($\Theta_\mathrm{RSG}$) and AGB ($\Theta_\mathrm{AGB}$) visibilities (computed from from the RHD simulations of Table~\ref{simus}). The arrows in the first column and the horizontal dashed lines indicate the wavelength interval where the telluric spectrum at the observing site is weak (Fig.~\ref{example_water}).}
 \label{angulardiameter}
 \begin{tabular}{c|cccc}
 Wavelength [$\mu$m] & $\Theta_\mathrm{UD}$ & $\Theta_\mathrm{RSG}$ & $\Theta_\mathrm{AGB}$ \\
                   & [mas]                              &                                [mas] &  [mas] \\
\hline
    & & $L$ and $M$ bands           & \\
    \hline
    \hdashline
2.92 $\Downarrow$ & 26.16 $\pm$ 0.17 & 26.83 $\pm$        0.29 &       20.29 $\pm$      0.22 \\
2.99 & 26.21 $\pm$ 0.14 & 26.77 $\pm$    0.25 &       20.13 $\pm$      0.18 \\
3.05 & 26.38 $\pm$ 0.14 & 27.10 $\pm$    0.26 &       20.86 $\pm$      0.19 \\
3.11 & 25.44 $\pm$ 0.15 & 26.30 $\pm$    0.29 &       20.07 $\pm$      0.20 \\
3.17 & 24.09 $\pm$ 0.16 & 24.07 $\pm$    0.24 &       18.99 $\pm$      0.15 \\
3.23 & 21.94 $\pm$ 0.12 & 22.80 $\pm$    0.18 &       18.82 $\pm$      0.13 \\
3.29 & 21.36 $\pm$ 0.11 & 22.35 $\pm$    0.22 &       18.27 $\pm$      0.14 \\
3.35 $\Uparrow$ & 20.55 $\pm$ 0.12 & 21.22 $\pm$         0.20 &       18.31 $\pm$      0.14 \\
\hdashline
3.41 & 20.32 $\pm$ 0.13 & 21.27 $\pm$    0.21 &       18.61 $\pm$      0.16 \\
3.47 & 20.00 $\pm$ 0.13 & 21.09 $\pm$    0.22 &       19.35 $\pm$      0.17 \\
3.54 & 19.71 $\pm$ 0.13 & 21.05 $\pm$    0.23 &       19.84 $\pm$      0.19 \\
3.60 & 19.63 $\pm$ 0.14 & 20.99 $\pm$    0.24 &       20.14 $\pm$      0.21 \\
3.66 & 19.53 $\pm$ 0.14 & 21.18 $\pm$    0.25 &       20.40 $\pm$      0.28 \\
3.72 & 19.95 $\pm$ 0.16 & 21.32 $\pm$    0.29 &       20.25 $\pm$      0.30 \\
3.78 & 20.28 $\pm$ 0.18 & 21.52 $\pm$    0.33 &       21.50 $\pm$      0.35 \\
3.84 & 20.60 $\pm$ 0.20 & 21.83 $\pm$    0.36 &       22.55 $\pm$      0.39 \\
3.90 & 21.01 $\pm$ 0.21 & 21.91 $\pm$    0.38 &       22.48 $\pm$      0.40 \\
3.96 & 21.67 $\pm$ 0.22 & 22.92 $\pm$    0.39 &       23.96 $\pm$      0.42 \\
\hdashline
4.02 $\Downarrow$ & 22.73 $\pm$ 0.20 & 23.77 $\pm$       0.35 &       23.72 $\pm$      0.39 \\
4.08 & 23.87 $\pm$ 0.21 & 24.25 $\pm$    0.32 &       22.67 $\pm$      0.32 \\
4.14 & 25.29 $\pm$ 0.22 & 25.26 $\pm$    0.35 &       23.10 $\pm$      0.31 \\
4.20 & 27.02 $\pm$ 0.25 & 26.75 $\pm$    0.40 &       24.33 $\pm$      0.34 \\
4.27 & 28.80 $\pm$ 0.29 & 28.36 $\pm$    0.48 &       25.14 $\pm$      0.41 \\
4.33 & 31.84 $\pm$ 0.35 & 31.11 $\pm$    0.58 &       28.74 $\pm$      0.54 \\
4.39 & 34.10 $\pm$ 0.37 & 31.64 $\pm$    0.59 &       28.54 $\pm$      0.54 \\
4.45 & 38.55 $\pm$ 0.48 & 33.50 $\pm$    0.66 &       28.79 $\pm$      0.51 \\
4.51 & 38.06 $\pm$ 0.50 & 32.36 $\pm$    0.64 &       28.43 $\pm$      0.52 \\
4.57 & 38.65 $\pm$ 0.54 & 32.21 $\pm$    0.67 &       27.71 $\pm$      0.54 \\
4.63 & 41.38 $\pm$ 0.68 & 33.85 $\pm$    0.77 &       28.14 $\pm$      0.57 \\
4.69 $\Uparrow$ & 44.15 $\pm$ 0.98 & 28.88 $\pm$         0.56 &       25.73 $\pm$      0.55 \\    
\hdashline
\hline
     &                  & $N$ band           &                   \\
\hline
  8.10 & 133.66 $\pm$ 5.00 & -- & 91.87 $\pm$      6.55 \\
 8.40 & 149.59 $\pm$ 3.84 & -- & 103.64 $\pm$      5.19 \\
 8.70 & 164.17 $\pm$ 3.89 & -- & 116.32 $\pm$      5.63 \\
 9.00 & 179.36 $\pm$ 3.26 & -- & 128.73 $\pm$      4.78 \\
 \hdashline
 9.30 $\Downarrow$ & 191.71 $\pm$ 2.83 & -- & 143.71 $\pm$      4.66 \\
 9.59 & 202.89 $\pm$ 1.73 & -- & 153.91 $\pm$      3.54 \\
 9.89 $\Uparrow$& 210.49 $\pm$ 1.91 & -- & 157.80 $\pm$      3.73 \\
 \hdashline
10.19 & 215.67 $\pm$ 2.36 & -- & 171.12 $\pm$      4.89 \\
10.49 & 221.74 $\pm$ 2.85 & -- & 174.10 $\pm$      5.62 \\
10.79 & 226.60 $\pm$ 4.02 & -- & 173.66 $\pm$      6.84 \\
11.09 & 222.91 $\pm$ 5.00 & -- & 173.46 $\pm$      11.07 \\
11.39 & 230.87 $\pm$ 5.03 & -- & 178.95 $\pm$      9.12 \\
11.68 & 229.73 $\pm$ 4.68 & -- & 176.73 $\pm$      8.59 \\
11.98 & 221.32 $\pm$ 5.28 & -- & 173.06 $\pm$      10.96 \\
12.28 & 216.82 $\pm$ 5.53 & -- & 161.03 $\pm$      10.03 \\
12.58 & 221.34 $\pm$ 6.35 & -- & 161.99 $\pm$      10.10 \\
12.88 & 223.97 $\pm$ 7.87 & -- & 169.31 $\pm$      12.88 \\
\hline
 \end{tabular}
 \end{center}
\end{table}

\begin{figure}
   \centering
    \begin{tabular}{c}
     \includegraphics[width=1.\hsize]{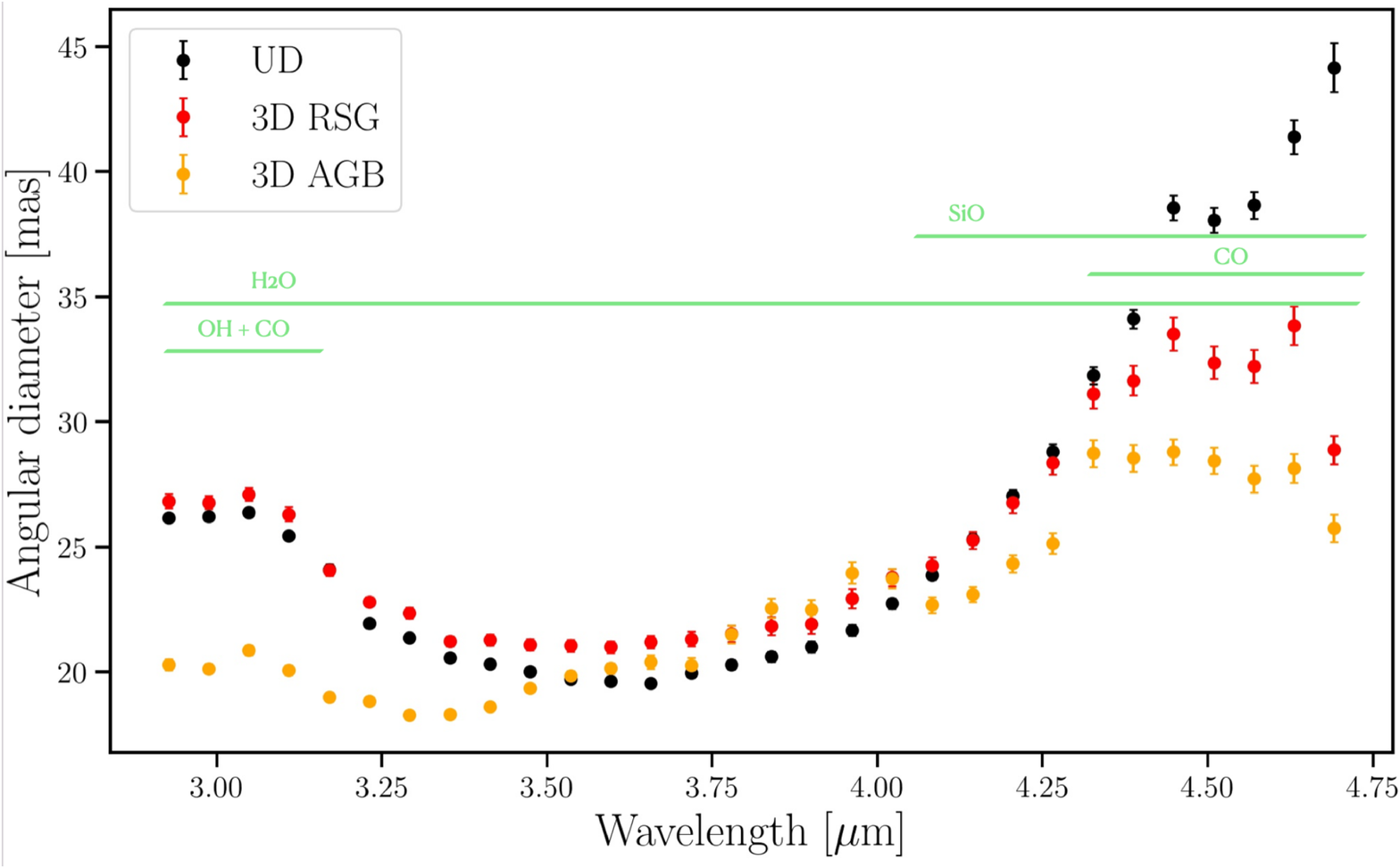}\\
     \includegraphics[width=1.\hsize]{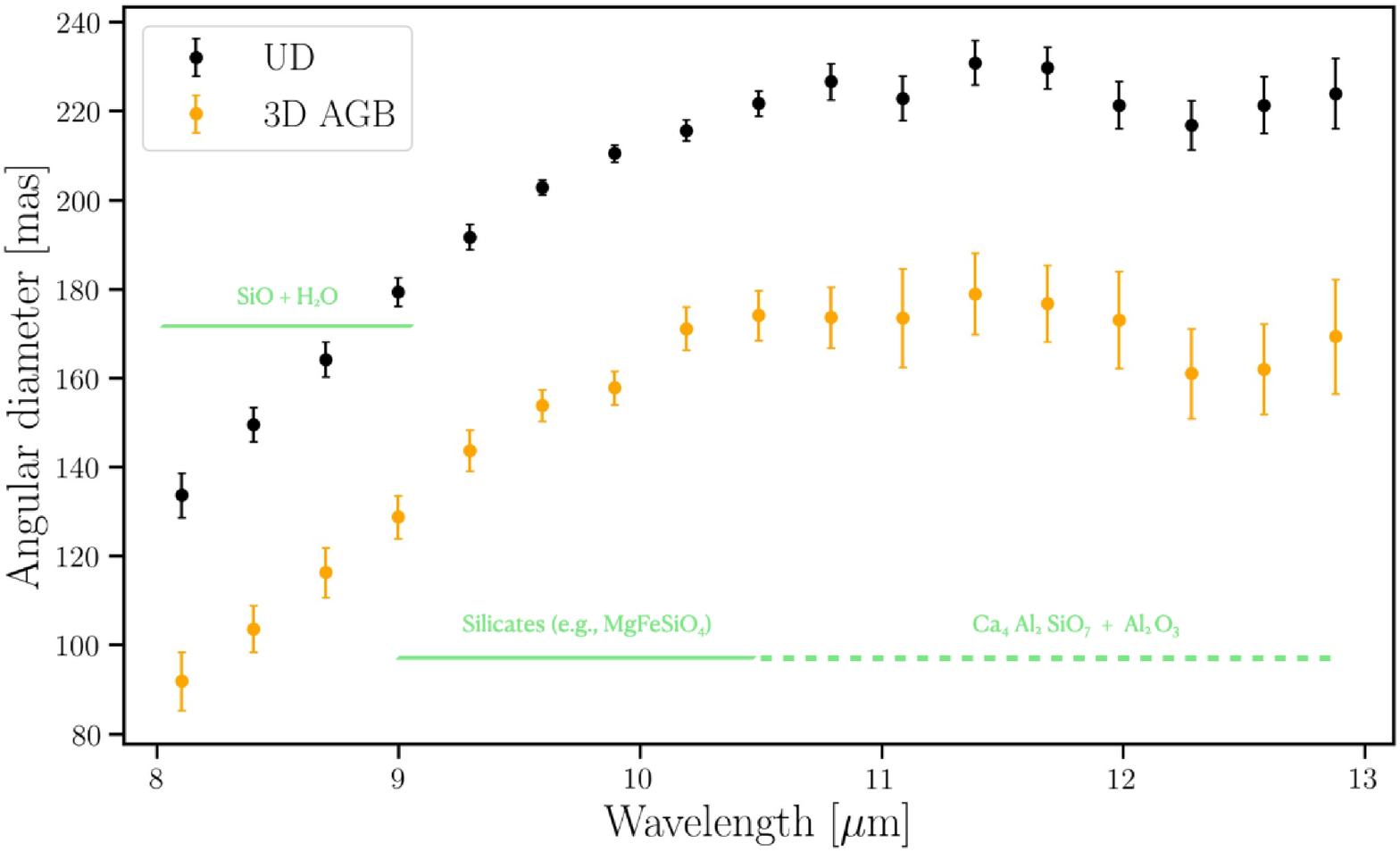}
     \end{tabular}
      \caption{Angular diameters measured (Table~\ref{angulardiameter}) with the RHD simulations of Table~\ref{simus} and a UD model in the $L$ and $M$ bands (\emph{top panel}) and the $N$ band (\emph{bottom panel}). The green horizontal lines highlight the principal chemical species.}
        \label{stellar_radius}
\end{figure}

Figure~\ref{stellar_radius} displays all the measured angular diameters of Table~\ref{angulardiameter} with highlighted in green the molecular and dust features that principally contribute to the observed flux. In the $L$ and $M$ bands, the apparent diameter observed depends on the opacity through the atmosphere. Theoretical work on AGB stellar atmospheres shows that, at low optical depth where molecules form, the dynamical pressure dominates over the gas pressure by factor of 5 to 10 \citep{2017A&A...600A.137F}. This makes possible the levitation of dense material into cool layers and allows the formation of irregular and non-spherical MOLsphere\footnote{As formulated for the first time by \cite{2000ApJ...538..801T}, it is a spherical region that is optically thin for particular molecular lines.} that, in turn, make the stellar size larger.\\ 
Figure~\ref{stellar_radius} (top panel) displays that the UD, RSG, and AGB diameters show similar behaviours with larger sizes at short wavelengths (where CO and OH are present), then smaller ones between $\sim$3.25 and $\sim$4.00\,$\mu$m, before increasing at longer wavelengths where SiO and CO dominate. In addition to these molecules, H$_2$O also contributes across the full spectral range. However, abrupt discontinuities are visible for RSG and UD diameters at (i) $\sim$3.15\,$\mu$m and (ii) $\sim$4.27\,$\mu$m. The strong variation in UD and RSG at $\sim$3.15\,$\mu$m indicates that these models are not adapted to describe the observed data since they underestimate the increase in size with the wavelength. This is not surprising because \cite{2015A&A...575A..50A} already pointed out that in 3D RSG simulations the dynamical pressure is not sufficient to enlarge the stellar atmosphere to the observed sizes in the $K$ band. On the other hand, the very extended atmospheres of the 3D AGB simulation shows a more regular increase of the radius from shorter to longer wavelengths, denoting that this simulation is more adapted to represent the data. It should be noted that the observed data in the spectral range between 
4.27 and 4.44\,$\mu$m are partially contaminated by the presence of noise (Fig.~\ref{example_obs}) that makes the determination of the visibility fit less precise, and, as a consequence, there is an abrupt change in the diameter values at precisely 4.27\,$\mu$m. 

In the $N$ band, Fig.~\ref{stellar_radius} (bottom panel) shows the increase of the angular diameter with the wavelength as already visible in Figs.~\ref{image2} and \ref{composite}. However, this result is purely indicative since either the UD or the AGB simulation lack dedicated dust opacities, where the dust features prevail.  %However, recently it should be noted that both global shocks induced by radial pulsations and small-scale shocks contribute to the levitation of material that  allows dust to form \citep{2019A&A...623A.158H}.

\subsection{Stellar surface convection in the $L$ and $M$ bands} 

\begin{figure}
   \centering
    \begin{tabular}{cc}
     \includegraphics[width=1.\hsize]{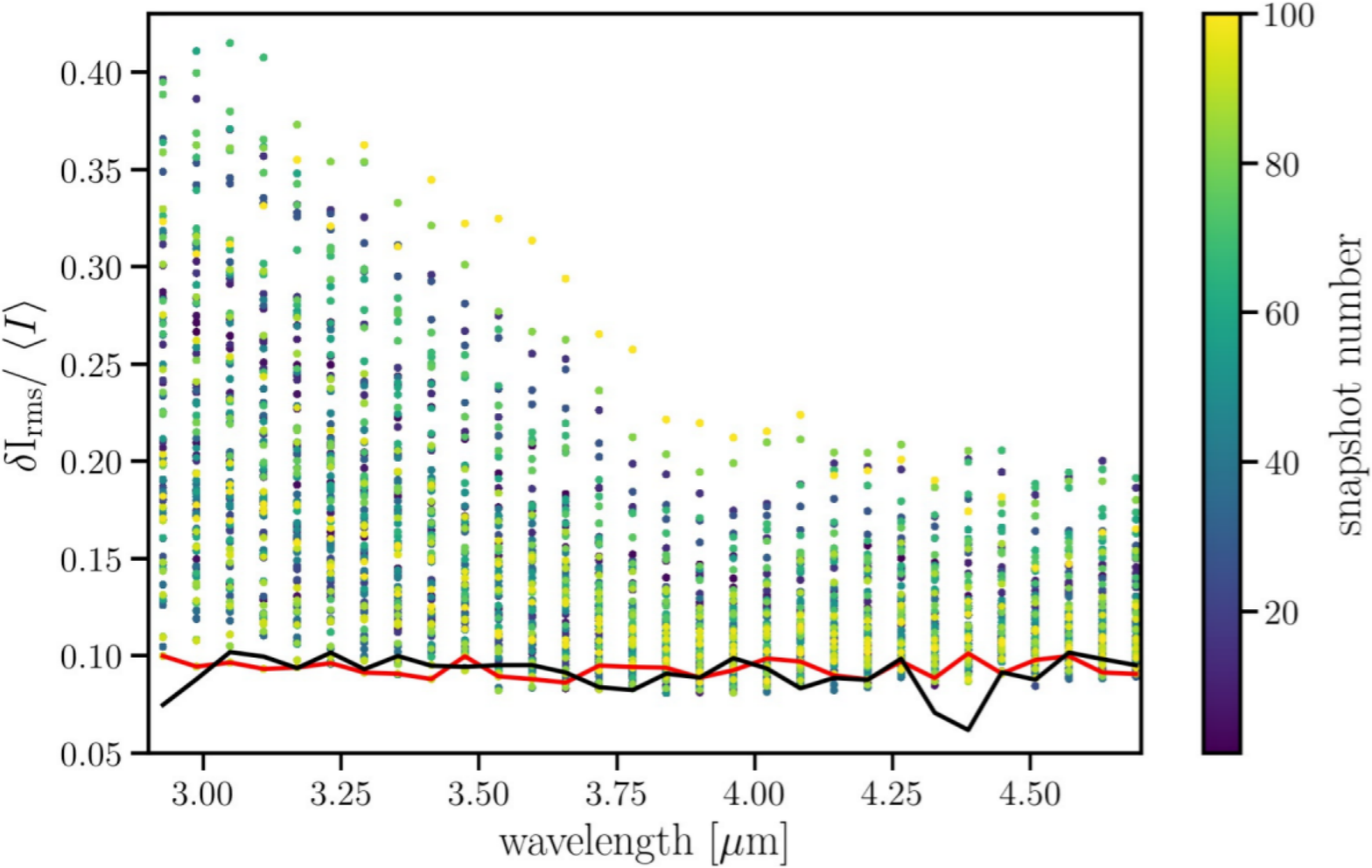} \\
     \includegraphics[width=1.\hsize]{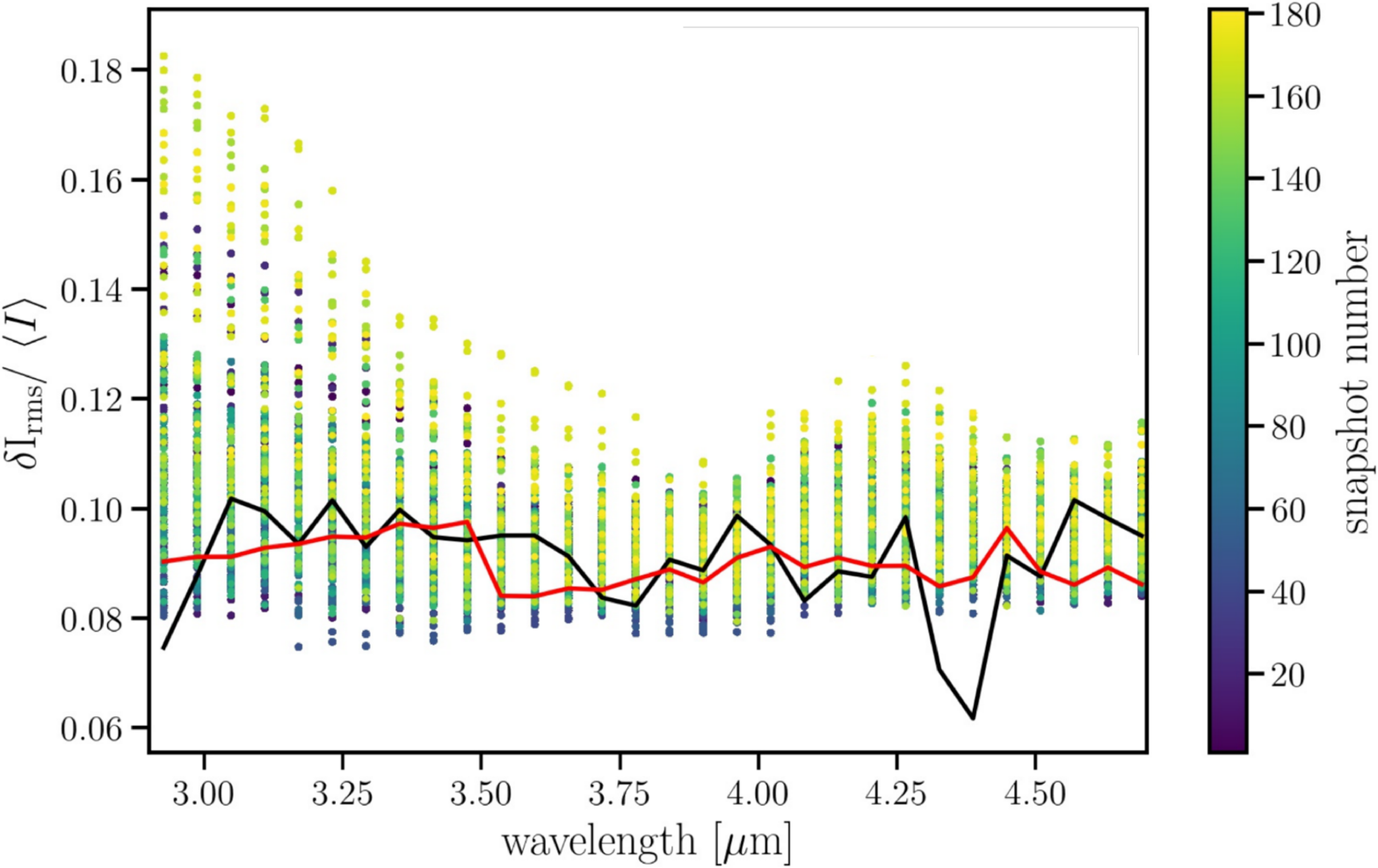} 
\end{tabular}
      \caption{Intensity contrast in the case of the AGB (\emph{top}) and the RSG (\emph{bottom}) simulations as a function of the wavelength in the $L$ and $M$ bands. Each coloured point indicates a different snapshot and the vertical bar on the right gives its number in the simulated temporal coverage. The black line displays the contrast of reconstructed images of Fig.~\ref{image1} and the red line the best matching snapshot across the wavelength bins.}
        \label{3D_comparison_rms}
\end{figure} 

In this section, we compare the stellar surface structure observed in the aperture synthesis imaging in the $L$ and $M$ bands to the RHD synthetic maps. The main aim is to focus the analysis on the central part of the stellar surface to point out evidence for the presence of convection-related structures.\\
We proceeded as follows: (i) We rebinned the 3D synthetic maps of both RHD simulations of Table~\ref{simus} to the resolution of 0.39\,mas pixel$^{-1}$ so that they effectively possess the same FOV and resolution as the reconstructed images (Fig.~\ref{image1}). (ii) We convolved them to the resolution of the interferometric beam (Fig.~\ref{uvplane}). (iii) We used the definition of \cite{2013A&A...557A...7T} to estimate the intensity contrast as $\delta I_{\mathrm{rms}}/\langle I\rangle$ for each snapshot, where $I$ is the intensity. This is performed only for the pixels within the stellar surface with a cut at 70$\%$ of the maximum of the ring-averaged intensity profile\footnote{A similar method based on the stellar radial cut is used in \cite{2020A&A...640A..23C} and \cite{2018A&A...614A..12M}.}. This procedure allows us to avoid outer areas dominated by the limb effect. (IV) We finally performed, for each snapshot and at a  given wavelength, a minimisation procedure based on the $\chi_{\mathrm{snapshot}}^2=\frac{\left|C_{\mathrm{observations}}-C_{\mathrm{3D}}\right|^2}{\sigma_{C_{\mathrm{observations}}}^2}$, where $C_{\mathrm{observations}}$ and $C_{\mathrm{3D}}$ are the intensity contrasts for the reconstructed image (observations) and the synthetic one (3D), while $\sigma_{C_{\mathrm{observations}}}$  is the standard deviation of the observed intensity contrast. Afterwards, we computed the reduced $\bar{\chi}^2$ as\\
    $\bar{\chi}^2$=$\frac{1}{N-1} \sum \chi_{snapshot}^2$ where $N$ is the total number of wavelength channels ($N=30$ in the $L$ and $M$ bands at low spectral resolution). In this way, our comparison is focused to matching simultaneously all the observed spectral channels with the same snapshot. This approach aims at having an additional wavelength dependent constraint for the simulation (i.e. the same snapshot should match all the wavelength channels). In the end, the best-matching snapshot is the one with the smallest $\bar{\chi}^2$.

\begin{figure*}
   \centering
    \begin{tabular}{cc}
     \includegraphics[width=1.\hsize]{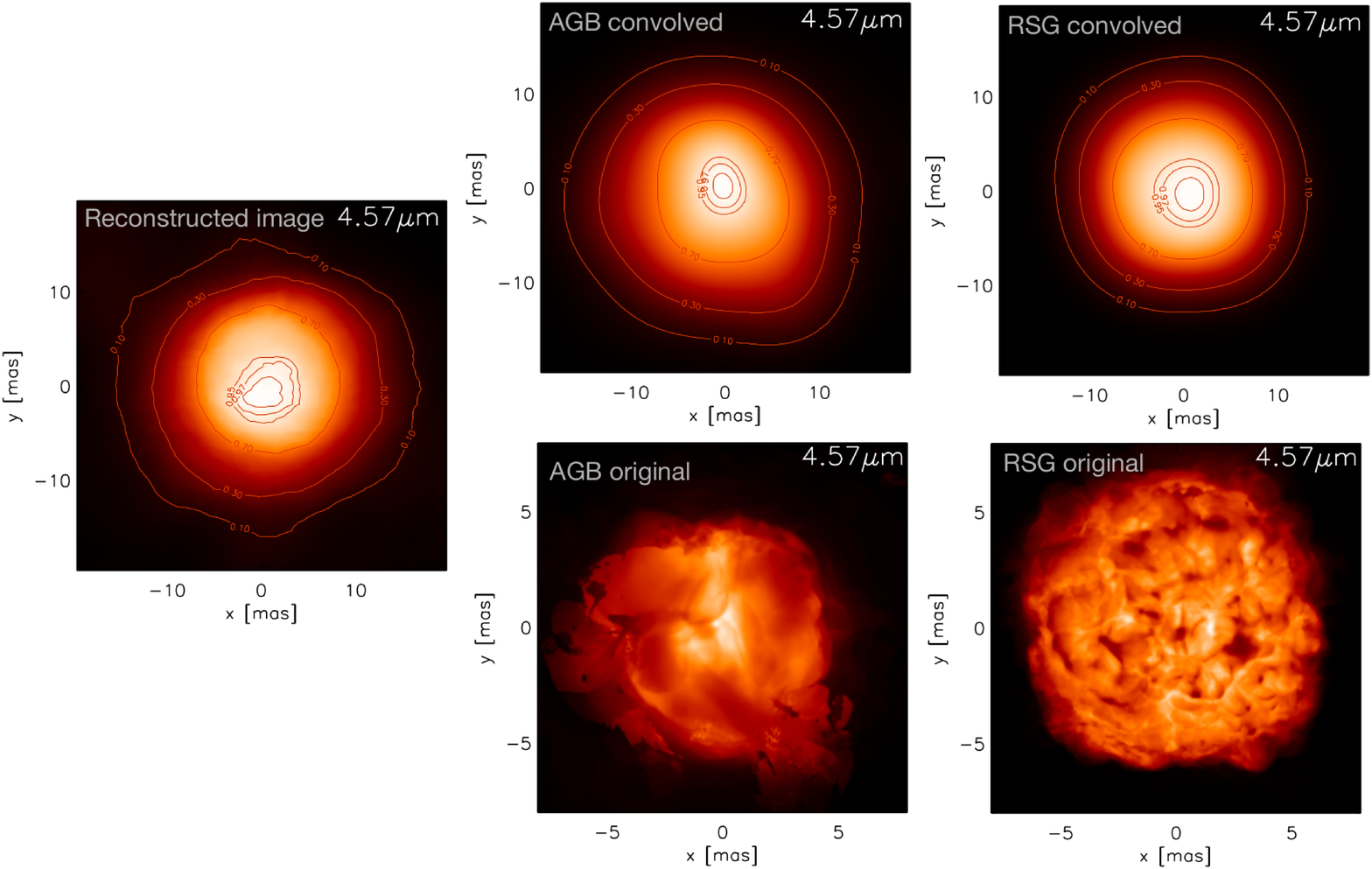}
\end{tabular}
      \caption{\emph{Top row:} Reconstructed image (left) and synthetic intensity maps computed for the best matching RHD simulations of Table~\ref{simus} at 4.57\,$\mu$m. The intensity is normalised between [0, 1]. The images, $256\times256$ pixels with 0.39\,mas pixel$^{-1}$, were convolved with the interferometric beam (Fig.~\ref{uvplane}). The contour lines are the same as in Fig.~\ref{image1}. East is left, while north is up. \emph{Bottom row:} Synthetic intensity maps from the above snapshots non-convolved with the interferometric beam. The peak intensity is normalised between [0, 1].
      The RSG snapshot $\bar{\chi}^2=$1.12, while the AGB one is $\bar{\chi}^2=$1.98.}
        \label{3D_comparison_images}
\end{figure*}

In the $L$ and $M$ bands, the best matching RSG snapshot has $\bar{\chi}^2=$1.12, and the has worst 29.80. For the AGB's best snapshot, $\bar{\chi}^2=$1.98 (worst 580.85). 
Fig.~\ref{3D_comparison_rms} shows the measured intensity contrast for the reconstructed and synthetic images. Fig.~\ref{3D_comparison_images} displays the best-matching snapshots for a particular wavelength channel (4.51\,$\mu$m) for the two considered RHD simulations of Table~\ref{simus}. 
In the $L$ and $M$ bands and restricting the comparison to the inner region only, the synthetic images of the AGB simulation are considerably wavelength dependent and, as a consequence, their morphology and photospheric extent noticeably change the apparent angular diameter (Fig.~\ref{angulardiameter}). In terms  of surface intensity contrast, their agreement with the reconstructed images is slightly worse in the case of the RSG simulation but the $\bar{\chi}^2$ values are very close. In the end, our 3D RHD simulations return a good agreement with the observations in terms of contrast of surface structures, meaning that they are adequate for interpreting the observed photospheric inhomogeneities as convection-related surface structures.

\section{N-band and tentative analysis with dust radiative transfer modelling}\label{radmodel}
        
\begin{figure}
   \centering
    \begin{tabular}{c}
     \includegraphics[width=0.95\hsize]{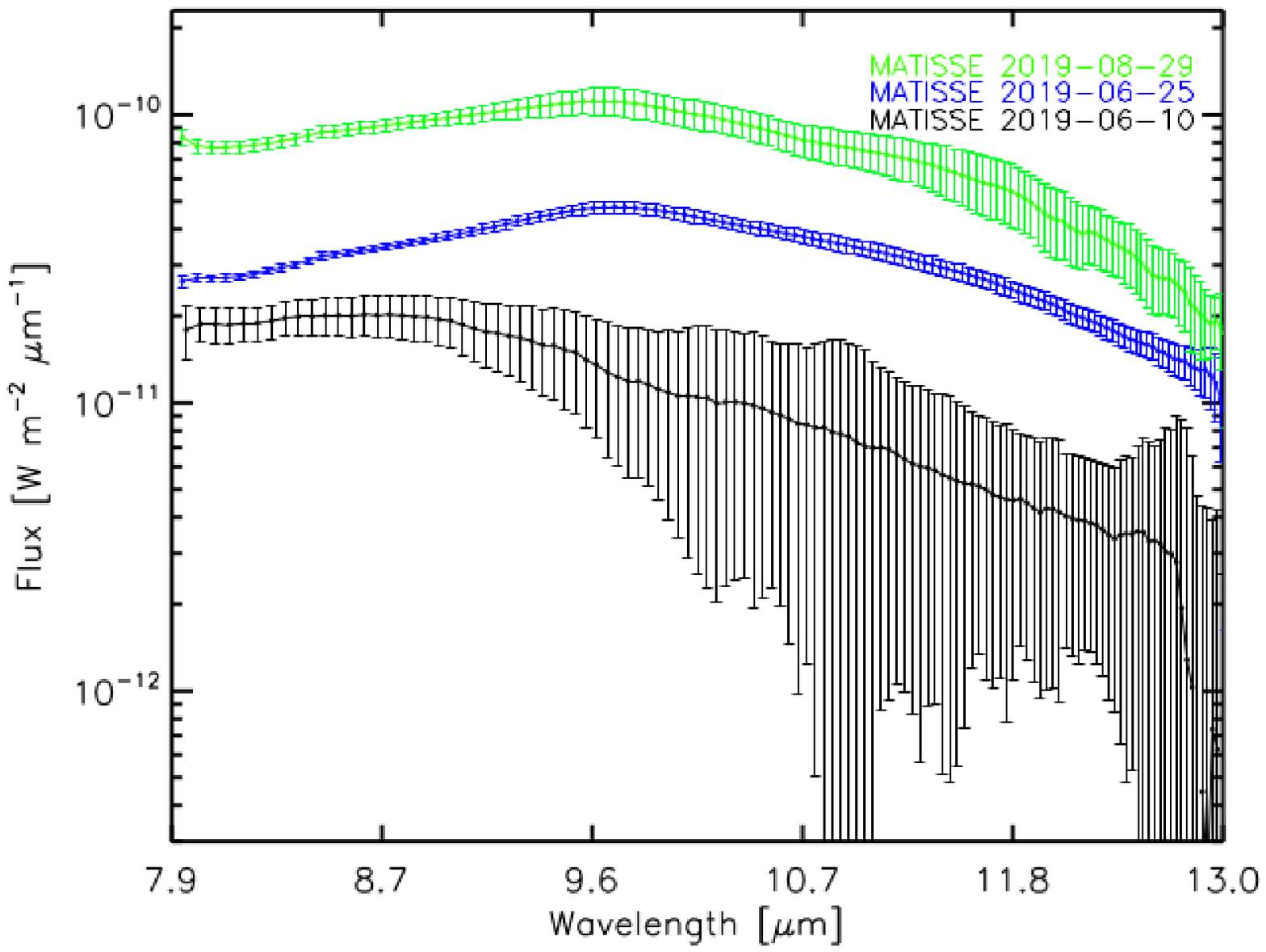}
\end{tabular}
      \caption{MATISSE calibrated fluxes for the three observing nights (Table~\ref{log}; see Sect.~\ref{datareductiontrue} for details for the details about calibration).}
        \label{matisse_flux}
\end{figure}

        We present in this section a qualitative analysis of the data obtained in the $N$ band. First, we used the MATISSE calibrated flux as described in Sect.~\ref{datareductiontrue}. Fig.~\ref{matisse_flux} displays the MATISSE calibrated flux. The calibrators for the night  2019-06-10 (see Table~\ref{log}) are too faint to allow a proper calibration and the data are extremely noisy. Both nights of 25 June 2019 and 28 June 2019 show similar shapes of the spectrum, but with lower quality for 2019-08-28. However, the mean absolute flux levels between these nights are different. Although there may be concerns about imperfect background subtraction and slight differences in air mass between the source and the calibrator (which are not corrected in the flux calibration process), the main cause of this discrepancy is the PHOENIX stellar synthetic spectra used for the calibrators. In our analysis, these spectra have been extrapolated from the near-IR to the $N$ band and, for the night of 2019-06-25, the theoretical spectrum of the calibrator underestimates the  measured SED by a factor of
two, while for the night 2019-08-28, there is an overestimation of the averaged flux measured. This partially explains the fact that the spectrum collected on 2019-06-25 is weaker than the one from 2019-08-28.

In conclusion, the accuracy of our flux calibration process does not allow us to obtain a consistent absolute calibration in terms of flux level between the different nights. For this reason, we focussed our analysis on interpreting only the overall shape of the spectrum. For this purpose, the nights of both 2019-06-25 and 2019-08-28 are used.         

        To model the observations in this wavelength range, where the dust contribution is significant, we used version 2.0 of the publicly available radiative transfer code \textsc{Radmc3D} \citep{2012ascl.soft02015D}. The code needs to specify a radiation source, and then, from a given input dust density distribution and dust opacities, it first computes the dust temperature distribution \citep[dust continuum radiative transfer, ][]{2001ApJ...554..615B}. Then, at every chosen wavelength, a scattering Monte Carlo run is performed to compute the scattering source function, which is then followed by a ray-tracing to compute the SEDs and the images. We used a spherical grid centred on the star initially with 20 grid points in each direction, with an adaptive mesh refinement scheme: each cell is refined if the dust density ($\rho$) gradient $(\rho_\mathrm{max} - \rho_\mathrm{min})/\rho_\mathrm{max}>0.05$, $\rho_\mathrm{min}$ and $\rho_\mathrm{max}$ are the minimum and maximum density values of 50 random points taken into each cell, respectively. 
        For the stellar flux, we used the 1D hydrostatic synthetic spectra \citep{2007A&A...468..205L} for a star with effective temperature of 3700\,K and surface gravity equal to 0.0, as was proposed by \cite{2021Natur.594..365M}, who analysed the observations of the RSG $\alpha$~Ori. Beside the fact that \cite{2013A&A...554A..76A} found and effective temperature of 3700\,K for VX Sgr, we also tried other 1D spectra with effective temperature of 2900\,K and 3400\,K with worse results than the one at 3700\,K.\\
        To model the apparent extension beyond the stellar radius, we used an envelope surrounding the star with three dust species: melilite \citep[Ca$_2$Al$_2$SiO$_7$, ][]{1998A&A...333..188M}, corundum \citep[Al$_2$O$_3$][]{2013A&A...553A..81Z}, and olivine \citep[MgFeSiO$_4$][]{1994A&A...292..641J,1995A&A...300..503D}. The choice of these species is based of the precedent findings for RSG stars \citep{2009A&A...498..127V}. The optical constants were obtained from the Jena database\footnote{\url{https://www.astro.uni-jena.de/Laboratory/OCDB/}}. We set the stellar angular diameter to 8.82 mas \citep[the near infrared value derived by Chiavassa et al. 2010]{2010A&A...511A..51C} to scale the synthetic stellar flux. To account for the extension of the object across the $N$ band (Fig.~\ref{image2} and Table~\ref{angulardiameter}), we computed SEDs with a flux contribution coming from a diameter of 130 mas. 
        
\begin{figure}
   \centering
    \begin{tabular}{c}
     \includegraphics[width=0.95\hsize]{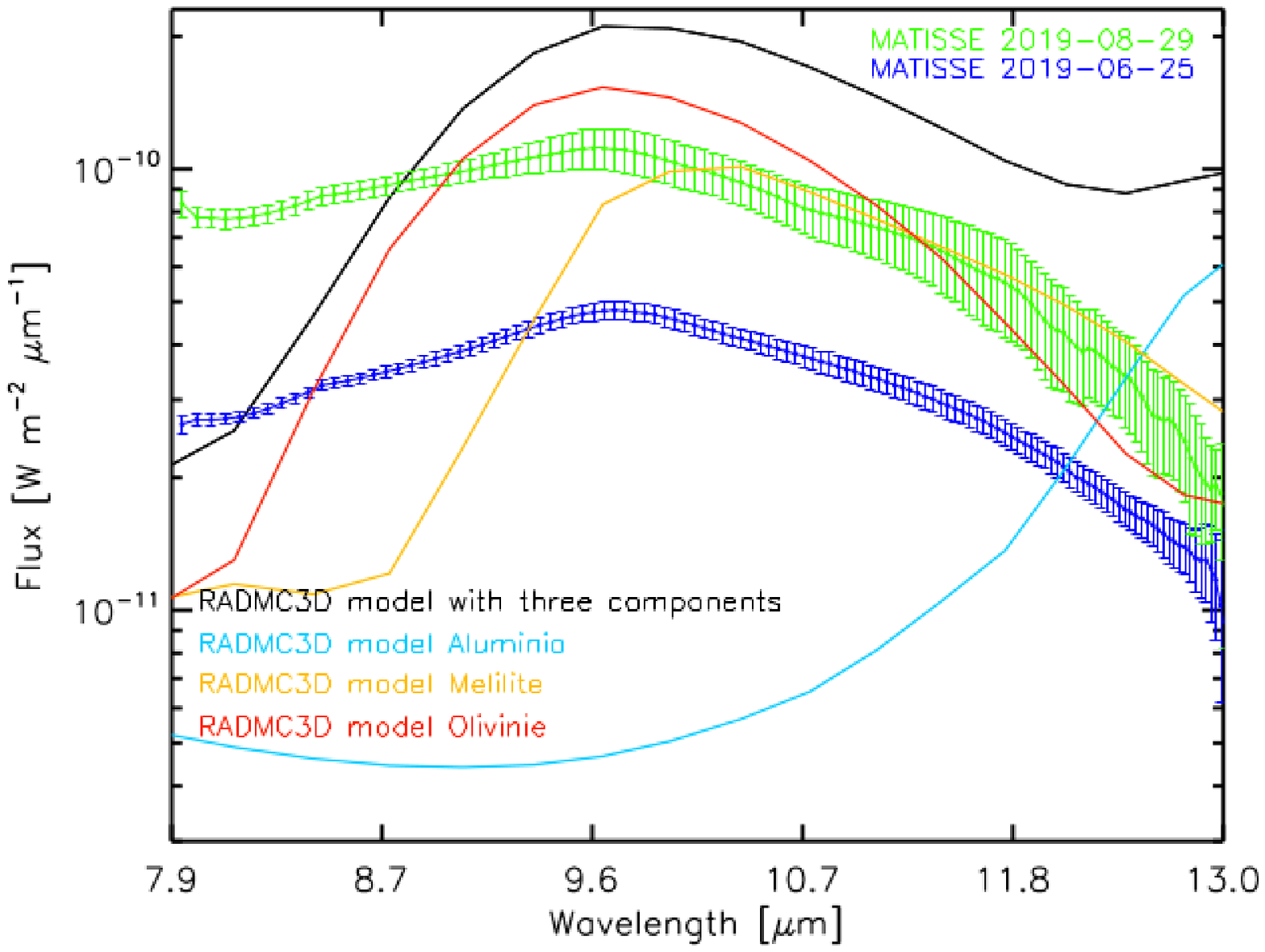}\\
      \includegraphics[width=0.95\hsize]{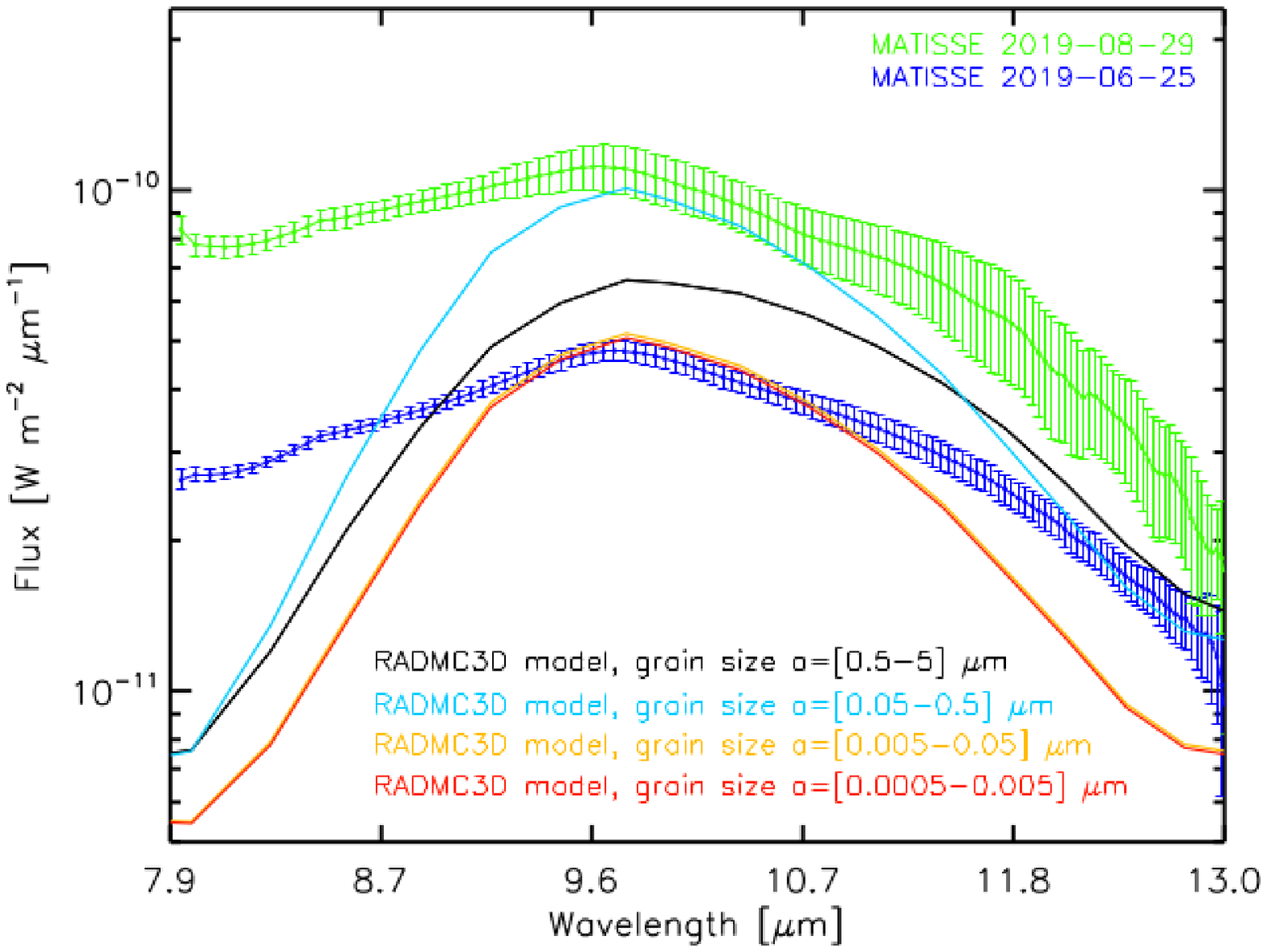}
\end{tabular}
      \caption{\emph{Top panel:} SEDs computed with \textsc{Radmc3D} compared to MATISSE calibrated flux from Fig.~\ref{matisse_flux}. The black curve shows the model with all three dust species included (melilite, corundum, and olivine), while the different colours (red, orange, and light blue) show one species at a time. \emph{Bottom panel:} SEDs computed with \textsc{Radmc3D} for olivine only and different grain size ranges.}
        \label{sed_fitting}
\end{figure}

Figure~\ref{sed_fitting} (top) displays an example where we set a grain size, $a$, ranging between [0.5-5]$\, \mu$m, with a size distribution of $n(a)\propto a^{-3.5}$ \citep{1977ApJ...217..425M}, for all the three species. We set the inner edge of the dust density distribution at 3.5 $R_\star$ to match the reconstructed images in Fig.~\ref{image2}. The dust mass-loss rate is set at $2\times 10^{-7}$\,$M_\odot$ yr$^{-1}$ for each of the three species. Moreover, Fig.~\ref{sed_fitting} (bottom) shows several \textsc{Radmc3D} for the olivine with different grain size ranges. The olivine flux decreases when reducing the grain size range and this behaviour is similar for all the other species. The general trend of the MATISSE data, for wavelength longer than $\sim$8.5\,$\mu$m, is in qualitative agreement with the presence of olivine and melilite, while corundum does not seem to be visible. At shorter wavelengths, a further contamination of other molecules not included here (e.g. SiO) may cause the discrepancy. Unfortunately, it is not possible to firmly conclude on the interpretation of the current data because of the difficulty in constraining the parameters of \textsc{Radmc3D} models using the limited accuracy of the absolute flux calibration.

\section{Conclusions}

Our MATISSE observations unveil the photospheric extent for the first time, as well as the circumstellar environment of VX~Sgr across the $L$,  $M$, and $N$ bands. We reconstructed monochromatic images using the MIRA software. They  show a complex morphology with brighter areas, whose characteristics depend on the wavelength probed. In the $N$ band, the central object is still visible at shorter wavelengths, but then the circumstellar brightness grows at longer wavelengths from 10 up to 40$\%$ of the central region.

We employed 3D RHD simulations of AGB and RSG stars, carried out with CO$^5$BOLD,
to calculate synthetic intensity maps in the same observed spectral channels of the MATISSE observations. As a first step, we measured VX~Sgr monchromatic angular diameters using a UD model fit and the azimuthally averaged intensity profiles from AGB and RSG simulations applied to the MATISSE interferometric data. The apparent diameter measured in the $L$ and $M$ bands depends on the opacity through the atmosphere: VX~Sgr features larger sizes at short wavelengths (where CO and OH are present), then it becomes smaller between $\sim$3.25 and $\sim$4.00\,$\mu$m, before increasing at longer wavelengths where SiO and CO dominate. On top of these molecules, H$_2$O also contributes across the full spectral range. 
The dynamical pressure dominates over the gas pressure at low optical depth, where molecules form \citep{2017A&A...600A.137F}. As a consequence, it makes the levitation of material possible that in turn allows the formation of an irregular and non-spherical MOLsphere. This extended photosphere is what is observed by MATISSE. In this context, the comparison with the $L$ and $M$ images denoted also that the photospheric extent predicted in 3D RSG simulation is not sufficient, as already shown in the K band \citep{2015A&A...575A..50A}.

The spectrum in the $N$ band may be characterised by the signature of several dust features \citep[][]{2017A&A...600A.136P} with a consequence on the brightness and on the general morphology of the star and its circumstellar environment. The angular diameter increases with wavelength. However, our comparison is purely indicative because, where the dust features prevail, the AGB simulation we used does not include source terms or dedicated opacities for dust. 

We also compared the stellar surface structure observed in the reconstructed images in the $L$ and $M$ bands to RHD synthetic maps. Our 3D RHD simulations return a good agreement with the observations in terms of contrast of surface structures. We conclude that the photospheric inhomogeneities observed in the reconstructed images could be interpreted as convection-related surface structures.\\

        To reproduce the observations in the $N$ band, we used the publicly available radiative transfer code \textsc{Radmc3D} to calculate the SED of an envelope surrounding the star with three dust species: melilite (Ca$_2$Al$_2$SiO$_7$), corundum (Al$_2$O$_3$), and olivine (MgFeSiO$_4$). The general trend of MATISSE data is in qualitative agreement with the presence of melilite and olivine but not corundum. Nevertheless, it is not possible to firmly conclude on the interpretation of the current data because of the difficulty in contrasting the model parameters using the limited accuracy of our absolute flux calibration.
        
Even if the MATISSE observations are extremely useful to characterise and resolve the stellar surface as well as its circumstellar environment, they are not suitable to firmly conclude on the nature of VX~Sgr. The reason is twofold: on one hand, the spectral resolving power used here ($R=$35) is not sufficient to observe the spectral lines in detail, which could provide crucial information on the velocity fields and atmospheric stratification of the observed object \citep{2011A&A...535A..22C}; on the other hand, RSG and AGB stars display similarities in interferometric observables \citep[e.g. ][]{2010A&A...511A..51C,2015A&A...575A..50A,2019Msngr.178...34W}, making their interpretation difficult. Time-dependent, spatially and spectroscopically resolved, multi-wavelength simultaneous observations will be key to probing the stellar parameters of VX~Sgr and characterising its photometric variability.

\begin{acknowledgements}
      This work was granted access to the HPC resources of Observatoire de la C\^ote  d'Azur $--$ M\'esocentre SIGAMM. This research has also benefited from the help of SUV\footnote{http://www.jmmc.fr/suv.htm}, the VLTI user support service of the Jean-Marie Mariotti Center\footnote{http://www.jmmc.fr}. Moreover, this work was supported by the "Programme National de Physique Stellaire" (PNPS) of CNRS/INSU co-funded by CEA and CNES. BF acknowledges funding from the European Research Council (ERC)
under the European Union's Horizon 2020 research and innovation programme Grant agreement No. 883867, project EXWINGS)
and the Swedish Research Council ({\it Vetenskapsr{\aa}det}, grant number 2019-04059).
The computations of 3D models were enabled by resources provided by the
Swedish National Infrastructure for Computing (SNIC) at UPPMAX. VH is supported by the National Science Center, Poland, Sonata BIS project 2018/30/E/ST9/00598. This research made use of  IPython, Numpy, Matplotlib, SciPy, and Astropy\footnote{Available at \url{http://www.astropy.org/}}, a community-developed core Python package for Astronomy \citep{2013A&A...558A..33A}. 
\end{acknowledgements}

\bibliographystyle{aa}
\bibliography{biblio.bib}

\begin{appendix} 

\section{Intensity RMS maps of reconstructed images}

\begin{figure*}
   \centering
    \begin{tabular}{c}
       \includegraphics[width=0.45\hsize]{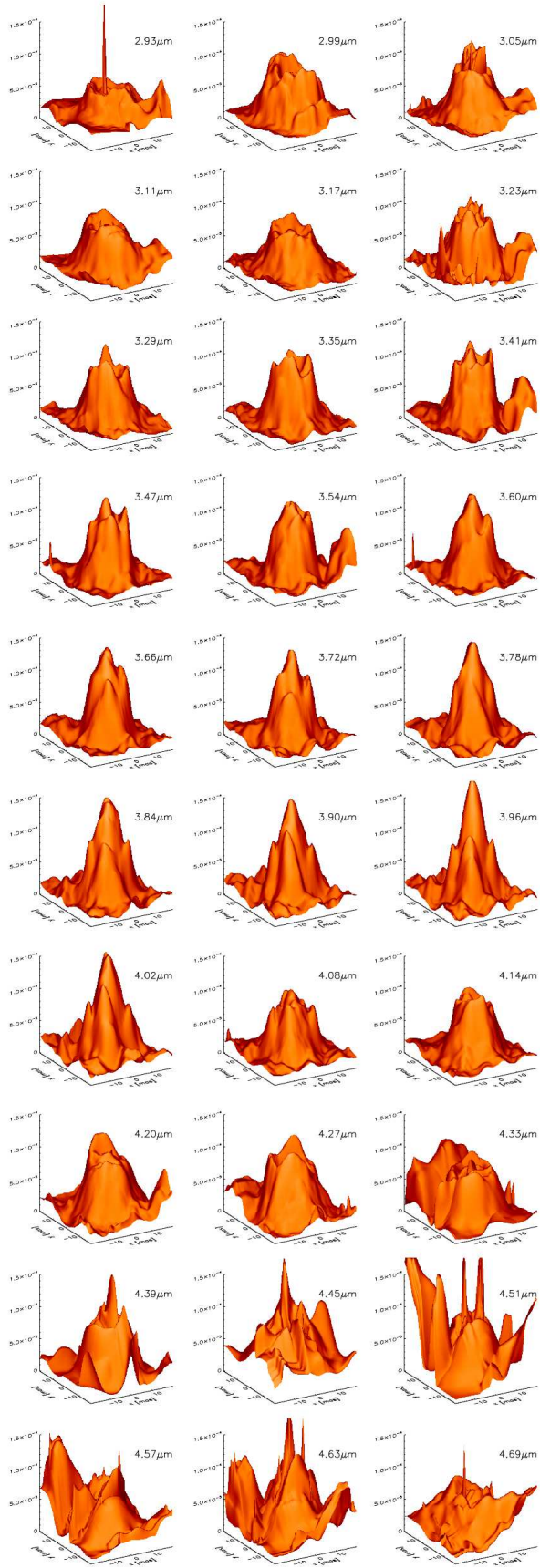} \\
    \end{tabular}
      \caption{Intensity RMS of the reconstructed images in the $L$ and $M$ bands from Fig.~\ref{image1}. As reported in Sect.~\ref{datareductiontrue}, the data quality of the images at wavelengths lower than $\sim$3.3 $\mu$m and between 4.0 and 4.7 $\mu$m, is compromised by the telluric spectrum at the observing site. The z-axis scale is the same for all the panels.}
        \label{RMSimage1}
           \end{figure*}

       \begin{figure*}
   \centering
    \begin{tabular}{c}
       \includegraphics[width=0.75\hsize]{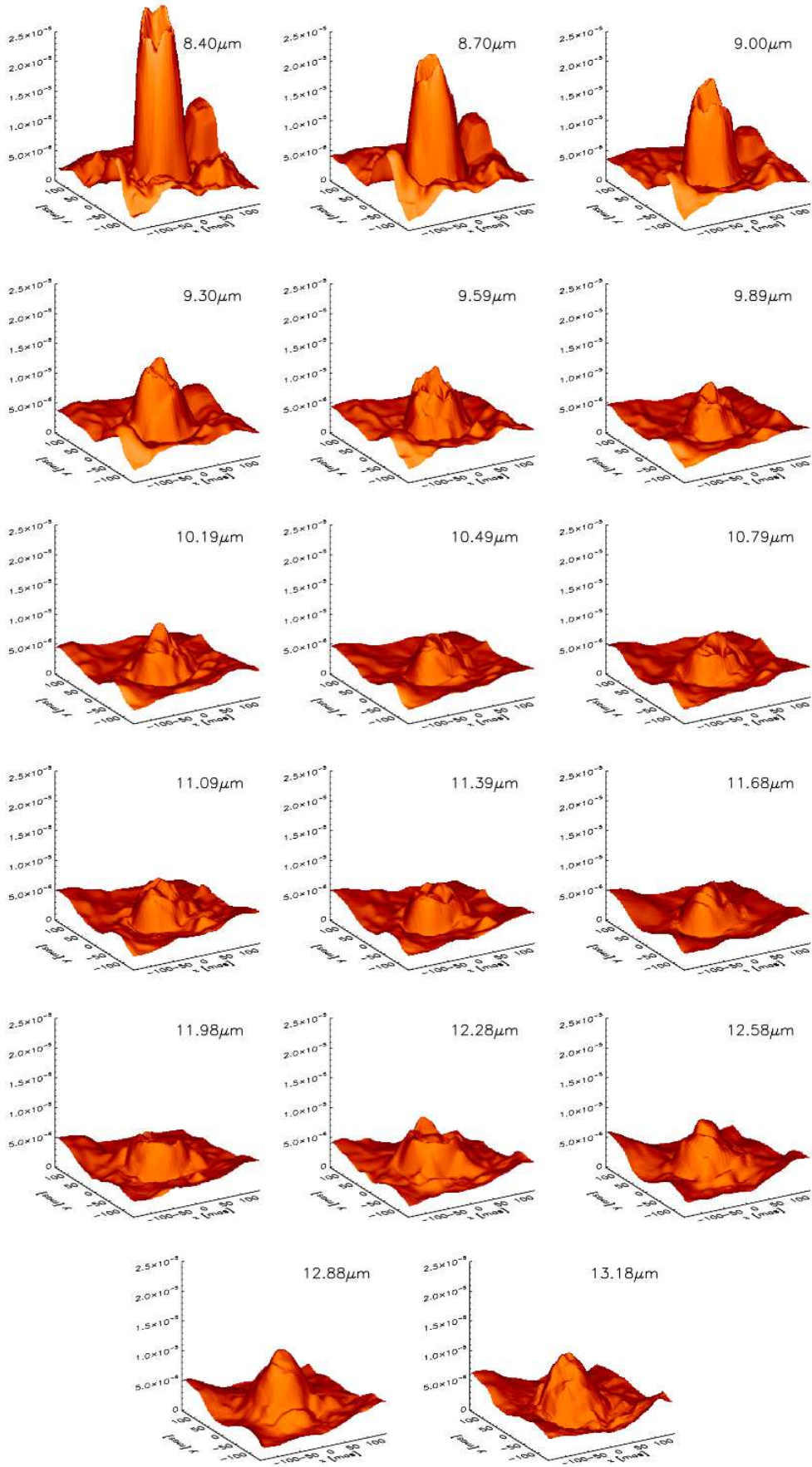} \\
    \end{tabular}
      \caption{Same as Fig.~\ref{RMSimage1}, but for the N band. The telluric spectrum dominates between 9.3 and 9.9 $\mu$m.}
        \label{RMSimage2}
           \end{figure*}      
           
\end{appendix}


\end{document}